\documentclass[twocolumn,showpacs,aps,prl,superscriptaddress]{revtex4}

\usepackage{graphicx} 
\usepackage{dcolumn} 
\usepackage{amsmath} 
\usepackage{epsfig} 
 
\usepackage{hhline} 
\usepackage{soul} 
\usepackage{tabularx,pifont,multirow} 
\usepackage{colordvi} 
 
\input babarsym 
\def\CPV {\ensuremath{C\!PV}\xspace}

\def\x{\ensuremath{x}\xspace} 
\def\y{\ensuremath{y}\xspace}

\def\Lik  {\ensuremath{\cal L}\xspace} 
\def\twoDLL  {\ensuremath{-2\Delta\ln{\cal L}}\xspace} 
\def\D  {\ensuremath{D}\xspace} 
\def\K  {\ensuremath{K}\xspace} 
\def\bea{\begin{eqnarray}} 
\def\eea{\end{eqnarray}} 
\def\nn{\nonumber} 
\def\mdz{\ensuremath{m_{\Dz}}\xspace}

\def\deltam{\ensuremath{\Delta m}\xspace} 

\def\t{\ensuremath{t}\xspace} 
\def\sigmat{\ensuremath{\sigma_t}\xspace} 
 
\def\tzero{\ensuremath{t_0}\xspace}

\def\splus{\ensuremath{s_{\rm +}}\xspace} 
\def\sminus{\ensuremath{s_{\rm -}}\xspace} 
\def\szero{\ensuremath{s_{\rm 0}}\xspace}

\def\pipsoft   {\ensuremath{\pi^{+}_{s}}\xspace} 
 
\def\pisoft   {\ensuremath{\pi_{s}}\xspace} 
 
\def\hp{\ensuremath{h^{+}}\xspace} 
\def\hm{\ensuremath{h^{-}}\xspace} 
\def\kshh{\ensuremath{\KS h^{+} h^{-}}\xspace} 
\def\kspipi{\ensuremath{\KS \pip \pim}\xspace} 
\def\Dztokspipi{\ensuremath{\Dz \to \kspipi}\xspace} 
\def\kskk{\ensuremath{\KS \Kp \Km}\xspace} 
\def\Dztokskk{\ensuremath{\Dz \to \kskk}\xspace} 
 
\newcommand{\BABARPubYear}    {10} 
\newcommand{\BABARPubNumber}  {004} 
 
\newcommand{\SLACPubNumber} {14087}

\def\figurebox#1#2#3{%
    \def\arg{#3}%
    \ifx\arg\empty 
    {\hfill\vbox{\hsize#2\hrule\hbox to #2{\vrule\hfill\vbox to #1{\hsize#2\vfill}\vrule}\hrule}\hfill}%
    \else 
    {\hfill\epsfbox{#3}\hfill}%
    \fi} 
 
\def\beq{\begin{equation}} 
\def\eeq{\end{equation}} 
\def\bea{\begin{eqnarray}} 
\def\eea{\end{eqnarray}} 
\def\bq{\begin{quote}} 
\def\eq{\end{quote}} 
\def\ben{\begin{enumerate}} 
\def\een{\end{enumerate}} 
\def\nn{\nonumber}

\newcommand{\phm}{\ensuremath{\phantom{-}}} 
 
\newcommand{\phz}{\ensuremath{\phantom{0}}} 
 
\newcommand{\phdag}{\ensuremath{\phantom{\dagger}}} 
\def\kspipi{\ensuremath{\KS \pip \pim}\xspace} 
 
\def\kskk{\ensuremath{\KS \Kp \Km}\xspace} 
 
\def\kshh{\ensuremath{\KS h^+ h^-}\xspace} 

\def\Dztokspipi{\ensuremath{\Dz \to \kspipi}\xspace} 
 
\def\Dztokskk{\ensuremath{\Dz \to \kskk}\xspace}

\def\D       {\ensuremath{D}\xspace} 
\def\K  {\ensuremath{K}\xspace} 
\def\t{\ensuremath{t}\xspace} 
\def\sigmat{\ensuremath{\sigma_t}\xspace}

\def\bea{\begin{eqnarray}} 
\def\eea{\end{eqnarray}} 
\def\nn{\nonumber}

\newcommand{\npa}       [1]  {\npBase\ A~{\bf #1}} 
 
\newcommand{\invgevc}{\ensuremath{{\mathrm{\,Ge\kern -0.1em V\!/}c}^{-1}}\xspace} 
\newcommand{\invmevc}{\ensuremath{{\mathrm{\,Me\kern -0.1em V\!/}c}^{-1}}\xspace}

\long\def\inst#1{\par\nobreak\kern 4pt\nobreak 
  {\it #1}\par\vskip 10pt plus 3pt minus 3pt} 
 
\begin{document}

\begin{flushleft} 
\babar-PUB-\BABARPubYear/\BABARPubNumber ~~~~~~~ SLAC-PUB-\SLACPubNumber   \\ 
\end{flushleft}

\title{ 
{ 
\large \bf \boldmath Measurement of \Dz-\Dzb mixing parameters using $\Dz \to \KS\pip\pim$ and $\Dz \to \KS\Kp\Km$ decays 
}
} 
 
%
\author{P.~del~Amo~Sanchez}
\author{J.~P.~Lees}
\author{V.~Poireau}
\author{E.~Prencipe}
\author{V.~Tisserand}
\affiliation{Laboratoire d'Annecy-le-Vieux de Physique des Particules (LAPP), Universit\'e de Savoie, CNRS/IN2P3,  F-74941 Annecy-Le-Vieux, France}
\author{J.~Garra~Tico}
\author{E.~Grauges}
\affiliation{Universitat de Barcelona, Facultat de Fisica, Departament ECM, E-08028 Barcelona, Spain }
\author{M.~Martinelli$^{ab}$}
\author{A.~Palano$^{ab}$ }
\author{M.~Pappagallo$^{ab}$ }
\affiliation{INFN Sezione di Bari$^{a}$; Dipartimento di Fisica, Universit\`a di Bari$^{b}$, I-70126 Bari, Italy }
% ...
\collaboration{The \babar\ Collaboration}
\noaffiliation

\date{\today}

\begin{abstract}  
\noindent 
We report a direct measurement of \Dz-\Dzb mixing parameters through 
a time-dependent amplitude analysis of the Dalitz plots of 
\Dztokspipi and, for the first time, \Dztokskk decays. 
The low-momentum pion $\pipsoft$ in the decay $\Dstarp \to \Dz\pipsoft$ identifies  
the flavor of the neutral $D$ meson at its production. 
Using 468.5 \invfb of \epem colliding-beam data  
recorded near $\sqrt s = 10.6~\gev$ by the \babar\ detector at the \pep2\  
asymmetric-energy collider at SLAC,  
we measure the mixing parameters  
$x= [1.6 \pm 2.3 \hbox{ (stat.)} \pm 1.2 \hbox{ (syst.)} \pm 0.8 \hbox{ (model)} ] \times10^{-3}$,  
and   
$y= [5.7 \pm 2.0 \hbox{ (stat.)} \pm 1.3 \hbox{ (syst.)} \pm 0.7 \hbox{ (model)} ]\times 10^{-3}$. 
These results provide the best measurement to date of $x$ and $y$.  
The knowledge of the value of $x$, in particular, is crucial for understanding the origin of mixing. 
\end{abstract}

\maketitle 
\indent 
Particle-antiparticle mixing and \CP violation (\CPV) in the  
charm sector are predicted to be very small 
in the standard model (SM)~\cite{ref:Petrov:2006nc,ref:Falk:2001hx,ref:Bigi:2000wn,ref:Wolfenstein:1985ft,ref:Falk:2004wg}.  
Evidence for \Dz-\Dzb mixing has been found only recently 
~\cite{ref:Aubert:2007wf,ref:Staric:2007dt,ref:Aaltonen:2007uc,ref:Aubert:2007en,ref:Aubert:2009ck,ref:Aubert:2008zh} 
and \CPV has not been observed. 
Although precise SM predictions for \Dz-\Dzb mixing  
are difficult to quantify,  
recent calculations of the mixing parameters \x  and \y allow for values as large  
as ${\cal O}(10^{-2})$~\cite{ref:Petrov:2006nc}. 
The analyses to date that have reported evidence for  
mixing 
have not been able to provide direct measurements of \x and \y.   
A time-dependent amplitude analysis of the Dalitz plot (DP) of \Dz mesons decaying  
into \kspipi and \kskk self-conjugate final states  
offers a unique way to access the  
mixing parameters \x and \y directly. 
In this Letter we study the time evolution of  
these three-body decays 
as a function of the position   
in the DP of squared invariant masses  
$\splus\!=\!m^2(\KS h^+)$, $\sminus\!=\!m^2(\KS h^-)$, where $h$ represents $\pi$ or $\K$, 
and report the most precise single measurements of \x and \y to date. 
The knowledge of the value of $x$, in particular, is crucial 
for understanding the origin of mixing and for 
determining whether contributions beyond the SM are present.  

We use the complete data sample of $468.5$~\invfb recorded near $\sqrt s = 10.6~\gev$  
by the \babar\ experiment~\cite{ref:Aubert:2001tu} at  the \pep2 asymmetric-energy \epem\ collider.  
The flavor of the neutral \D meson at production is identified through the charge of the $\pipsoft$ (``slow pion'') produced  
in the decay $\Dstarp \to \Dz \pipsoft$~\cite{ref:chargeconj}. 
The \Dz and \Dzb mesons evolve and decay as a mixture of the  
Hamiltonian eigenstates $D_1$ and $D_2$,  
with masses and widths $m_1$, $\Gamma_1$ and $m_2$, $\Gamma_2$, respectively.  
These mass eigenstates can then be written as linear combinations  
of flavor eigenstates,  
$| \D_{1,2}  \rangle\!=\!p | \Dz \rangle\!\pm\!q | \Dzb  \rangle$, where $|p|^2\!+\!|q|^2\!=\!1$. 
The mixing parameters are defined as $x = (m_1-m_2)/\Gamma$ and $y = (\Gamma_1-\Gamma_2)/2\Gamma$,  
where $\Gamma = (\Gamma_1+\Gamma_2)/2$ is the average decay width. 

Assuming no \CPV in the decay, the  
relation $\overline{{\cal A}}(\splus,\sminus)={\cal A}(\sminus,\splus)$ holds, where ${\cal A}$ and $\overline{{\cal A}}$  
are the decay amplitudes for a \Dz or a \Dzb into the final state \kshh  
as a function of the position in the DP. 
The time-dependent decay amplitude for a charm meson tagged at $\t=0$ as \Dz or \Dzb can then be written as  
\begin{eqnarray} 
&{\cal M}(\splus,\sminus,t) = {\cal A}(\splus,\sminus) g_{+}(t) +  
                    \frac{q}{p} {\cal A}(\sminus,\splus) g_{-}(t),\hspace{10pt}&  \nn \\ 
&{\cal\overline{M}}(\splus,\sminus,t) = \frac{q}{p}\overline{{\cal A}}(\splus,\sminus) g_{+}(t) +  
                     \overline{{\cal A}}(\sminus,\splus) g_{-}(t),\hspace{10pt}& \nn  
\label{eq:TimeDepAmp} 
\end{eqnarray} 
where $g_{\pm}(t)= 1/2\left[e^{-i(m_{1}-i\Gamma_{1}/2)t}\pm e^{-i(m_{2}-i\Gamma_{2}/2)t} \right]$ and 
$q/p=1$ if \CP is conserved in the mixing amplitude. 
The decay rates for \Dz and \Dzb are obtained by squaring ${\cal M}$ and ${\cal \overline{M}}$ respectively,  
and consist of a sum of terms depending on $(\splus, \sminus)$ and proportional  
to $\cosh(y \Gamma t)$, $\sinh( y\Gamma t)$, $\cos(x \Gamma t)$, and $\sin( x\Gamma t)$, 
all modulated by the exponential decay factor $e^{-\Gamma t}$. 
Assuming a  
model for  
${\cal A}(\splus,\sminus)$, 
it is possible to extract  
the mixing parameters \x and \y  
from the data, 
along with the amplitude model parameters and the proper-time resolution function. 
The variation of the distribution of the events in the DP as a function of the proper \Dz decay time 
is the signature of \Dz-\Dzb mixing. 
The sensitivity to \x and \y arises mostly from  
regions in the DP where Cabibbo-favored and  
doubly-Cabibbo-suppressed amplitudes interfere 
and from regions populated by \CP eigenstates~\cite{ref:babar_gamma_dalitz2008}.  
This method was pioneered by CLEO~\cite{ref:Asner:2005sz} and extended to a significantly larger  
data sample by Belle~\cite{ref:Abe:2007rd}. 

\indent 
The \Dz candidates are reconstructed in the \kspipi (\kskk) final state  
by combining \KS candidates with two oppositely-charged pions (kaons),  
with an 
invariant mass \mdz between 1.824 and 1.904 \gevcc. 
In order to reduce combinatorial background and to remove \Dz candidates from \B-meson decays, we require 
the momentum of the \Dz in the \epem center-of-mass frame to be greater than $2.5~\gevc$.   
The difference \deltam between the \Dstarp and \Dz reconstructed invariant  
masses is required to satisfy  
$0.143 < \deltam  < 0.149~\gevcc$. 
Each pion (kaon) track is identified using a likelihood particle identification algorithm based  
on \dedx ionization energy loss and Cherenkov angle measurements. 
The \KS candidates are selected by pairing two oppositely-charged pions  
whose invariant mass is within 9 \mevcc of the nominal  
\KS mass~\cite{ref:pdg2008}.  
We require the cosine of the angle between the \KS flight direction 
(defined by the \KS production and decay vertices) 
and the \KS momentum  
to be greater than 0.99, 
and a decay length projected along the \KS momentum to be greater than 10 times its error. 
These selection criteria suppress to a negligible level the 
background from $\Dz\to\pip\pim\hp\hm$ decays.  
For each charged track we  
require a transverse momentum with respect to the beam axis to be greater than $100~\mevc$, and  
for tracks from the \Dz decay we additionally require at least 
two hits in the two  
innermost 
layers of the silicon vertex tracker~\cite{ref:Aubert:2001tu}. 

\begin{figure}[!h] 
\begin{center} 
\begin{tabular} {cc}   
\includegraphics[height=4.1cm]{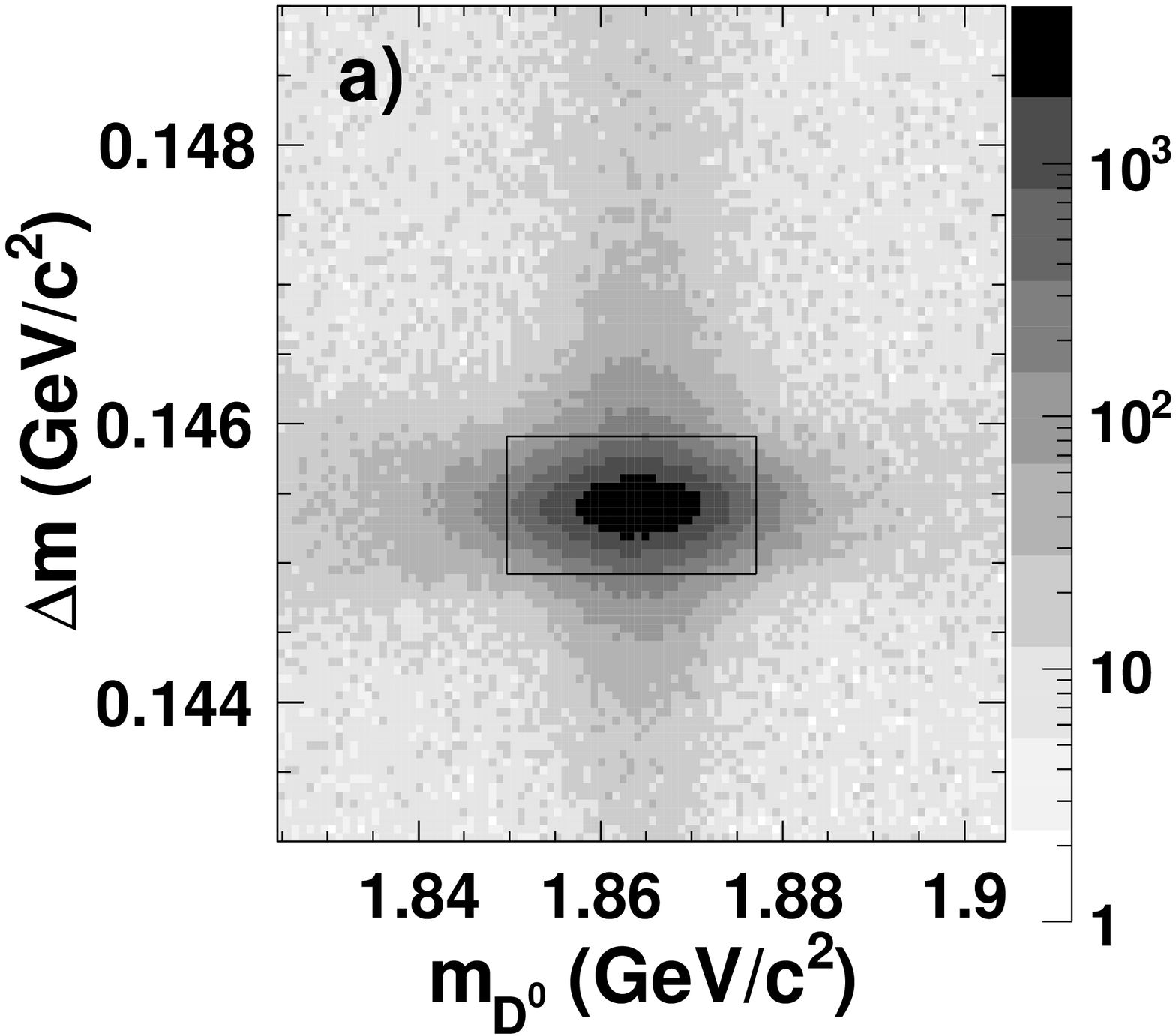} & 
\includegraphics[height=4.1cm]{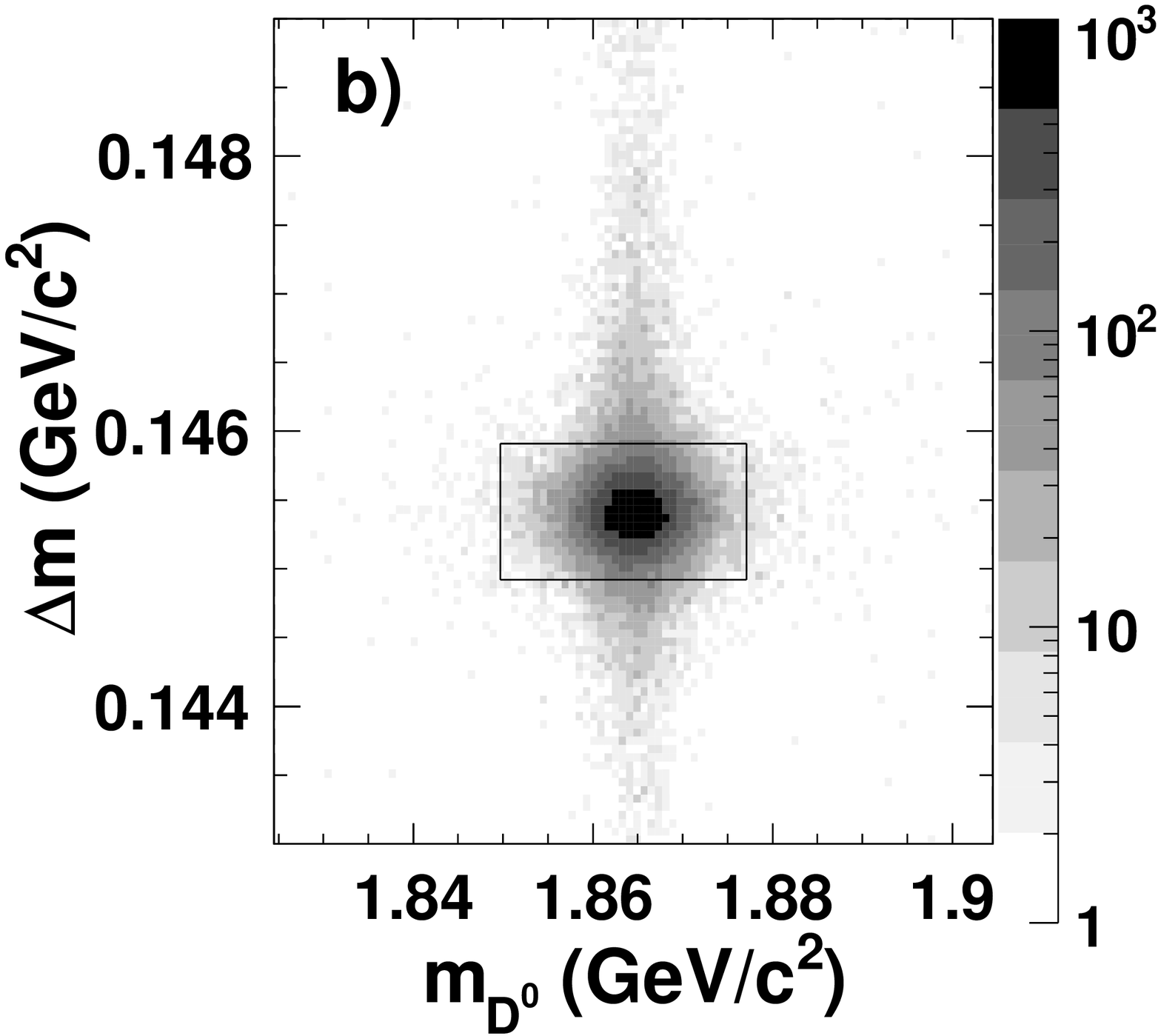} \\ 
\end{tabular}    
\caption{Distributions of \mdz-\deltam for (a) \kspipi and (b) \kskk  
data  
after all selection criteria.  
The gray scale indicates the number of events per bin.  
The rectangles enclose the signal region.  
} 
\label{fig:mD-DeltaM} 
\end{center} 
\end{figure}

The \Dz proper time $t$, and its error $\sigma_t$, are obtained through a kinematic fit  
of the entire decay chain which constrains the \KS and pion (kaon) tracks to originate  
from a common vertex and also requires the  
\Dz and the \pipsoft candidates to originate from a common vertex, constrained by the position and size  
of the \epem interaction region. 
This reduces the contribution from $\Dz\to\KS\KS$ decays (affecting only \kspipi)  
to $3\%$ of the total background. 
We retain candidates for which the $\chi^2$ probability of the fit  
is greater than 0.01\%,  
$|\t | < 6$~\ps, 
and $\sigmat<1$~\ps. 
The most probable value for \sigmat is about $0.2$ ($0.3$)~\ps for \kspipi (\kskk) signal candidates. 
For events where multiple \Dstarp candidates share one or  
more tracks, we keep the \Dstarp candidate with the highest  
$\chi^2$ probability. 
After applying all selection criteria, we find  
$744\hspace{2pt}000$ ($96\hspace{2pt}000$) \kspipi (\kskk) candidates. 
Their \mdz and \deltam distributions are shown in Fig.~\ref{fig:mD-DeltaM} and in~\cite{ref:epaps}.

\indent 
The mixing parameters \x and \y are determined from an unbinned, extended maximum-likelihood fit to the  
\kspipi and \kskk samples 
over the observables  
\mdz, \deltam, \splus, \sminus, \t, and \sigmat.  
First, the signal and background yields  
are determined from a fit to 
\mdz and \deltam distributions. 
For the subsequent fits, we restrict events to the signal region illustrated  
in Fig.~\ref{fig:mD-DeltaM}, defined to lie within twice 
the measured resolution around the mean \mdz and \deltam values, and holding the 
signal and background yields 
fixed to their signal window rescaled values. 
In the mixing fit, the \mdz and \deltam shapes are excluded to minimize correlations with  
the rest of the observables.  
Our reference fit allows for mixing but assumes no \CPV.  
We then allow for \CPV as a cross-check of  
the mixing results. 
To avoid potential  
bias,  
the mixing results were examined only  
after the fitting and analysis procedures were finalized. 

For each fit stage, different sub-samples are characterized separately:  
\kshh signal, random \pipsoft, misreconstructed \Dz,  
$\Dz\to\KS\KS$ events (for \kspipi decays only), 
and combinatorial background.  
The random $\pi_s^+$ component describes correctly reconstructed \Dz decays combined with a random slow pion. 
Misreconstructed \Dz events have one or more \Dz decay products either missing or reconstructed 
with the wrong particle hypothesis. 
We  account for $\KS\KS$ background events since they exhibit a  
characteristic DP distribution and a signal-like shape in the variables \t and \sigmat. 
Combinatorial background events are those not described by the above  
components.  
The functional forms of the \mdz, \deltam probability density functions (PDFs)  
for the signal and background components 
are chosen based on studies performed on large Monte Carlo (MC) samples.  
These account for the observed correlations between \deltam  
and \mdz for signal and misreconstructed \Dz events.  
The PDF parameters are determined from two-dimensional likelihood fits to data over the full \mdz and \deltam region, 
or over dedicated sideband data samples. 
We find  
$540\hspace{2pt}800\pm800$ ($79\hspace{2pt}900\pm300$) 
signal events  in the  
\kspipi (\kskk) signal region, with purities of $98.5\%$ ($99.2\%$),  
and reconstruction efficiencies of $14.4\%$ ($14.6\%$). 
Random \pipsoft, misreconstructed \Dz, and combinatorial background events 
account for $23\%$ ($53\%$), $52\%$ ($23\%$), and $22\%$ ($24\%$) of the total background. 
Projections of these fits and the contributions from the different background components  
can be found in~\cite{ref:epaps}.

The amplitudes ${\cal A}(\splus,\sminus)$  
are described by a coherent sum of quasi-two-body  
amplitudes~\cite{ref:babar_gamma_dalitz2008,ref:RevDalitzPlotFormalism}. 
The dynamical properties of the P- and D-wave amplitudes 
are parameterized through intermediate resonances with mass-dependent relativistic Breit-Wigner (BW) or Gounaris-Sakurai (GS) 
propagators, 
Blatt-Weisskopf centrifugal barrier factors, and 
Zemach tensors for the angular distribution~\cite{ref:RevDalitzPlotFormalism}. 
The $\pi\pi$ S-wave dynamics is described 
through a K-matrix formalism with the P-vector approximation and 5 poles~\cite{ref:AS,ref:babar_gamma_dalitz2008}.  
For the $\K\pi$ S-wave we include a BW for the $K^{*}_0(1430)^\pm$ with a  
coherent non-resonant contribution 
parametrized by a scattering length and effective range similar  
to those used to describe $\K\pi$ scattering data~\cite{ref:LASS,ref:epaps}. 
For the $\K\Kbar$ S-wave, a coupled-channel BW is used for the $a_0(980)$ isovector with 
BWs for the $f_0(1370)$ and $a_0(1450)$ states. 

We define PDFs to describe the dependence of the  
components in our  
event sample upon DP position $(\splus, \sminus)$ and upon decay time, \t.   
For signal, $\left|{\cal M}\right|^2$ or $\left|{\cal \overline{M}}\right|^2$  
is convolved with a proper-time resolution function different for  
\kspipi and \kskk events with  parameters determined by the  
mixing fit to the data. 
The resolution function is a sum of three Gaussians with one of  
the means allowed to differ from zero (the offset, $\tzero$), and two of the widths  
proportional to  
\sigmat.  
While $\tzero$ does not depend sensitively on the DP position,  
the ability to reconstruct \t varies as a function of $(\splus, \sminus)$. 
Hence, the observed distributions of \sigmat in a number of DP regions  
are included in the signal PDF. 
We apply corrections for efficiency variations and neglect the invariant mass resolution across the DP. 
The time-dependent PDF for the small random \pisoft background component is described by  
an equal combination of \Dz and \Dzb signal events assuming no mixing, since the slow pion  
is positive or negative with approximately equal probability, 
and  
carries little weight 
in the vertex fit.  
The PDFs for misreconstructed \Dz events and combinatorial background are determined 
from  \mdz sideband samples. 
A non-parametric approach is used to construct the DP distributions,  
while the proper-time distributions are described by a sum of two Gaussian  
functions,  
one of which has a power-law tail to account for a small long-lived component. 
The background components containing real and misreconstructed \Dz decays have different \sigmat distributions, which are 
determined from the signal and \mdz sideband regions. 
 
\begin{figure}[!h] 
\begin{center} 
\begin{tabular} {cc}   
\includegraphics[height=4.1cm]{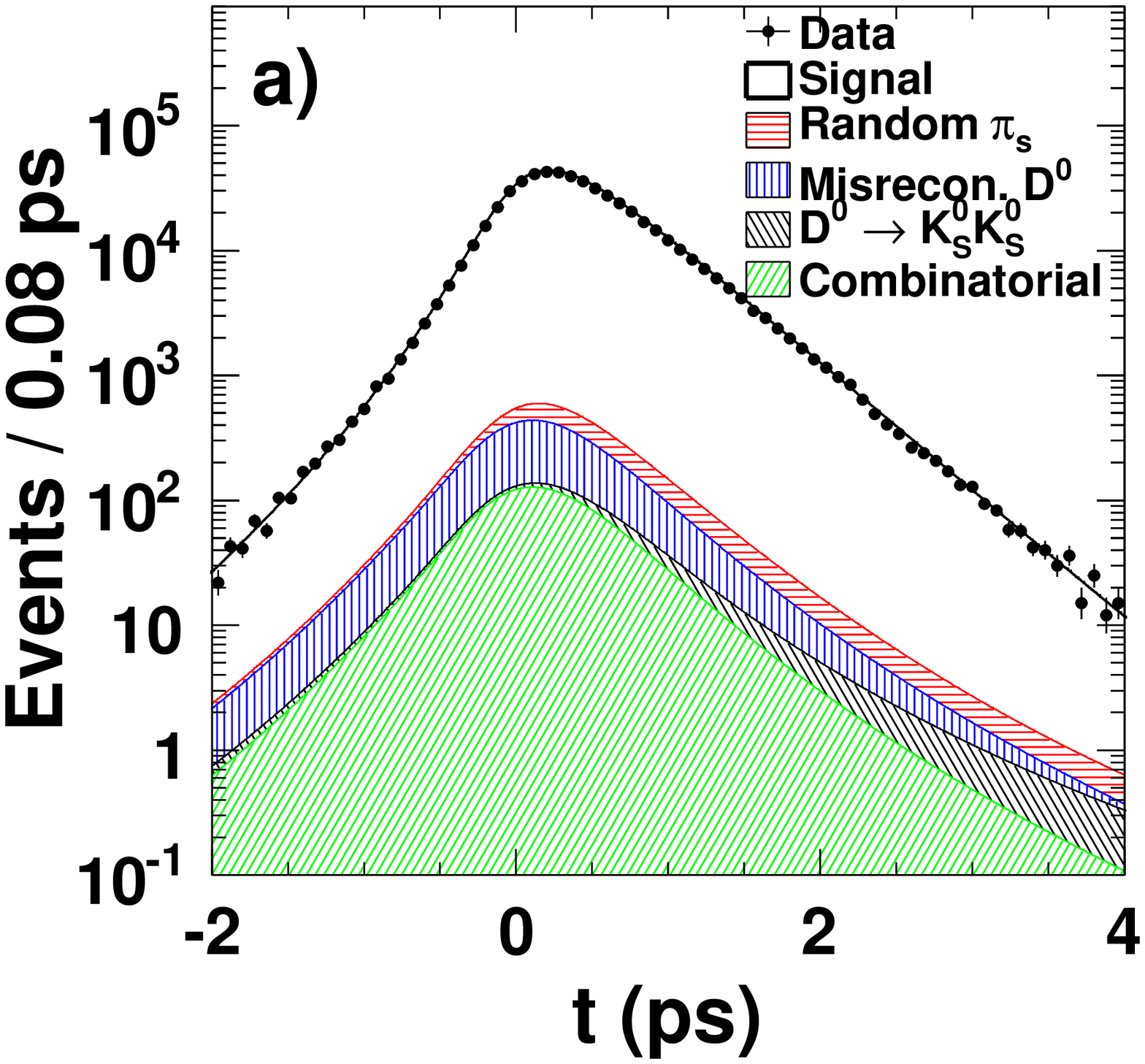} & 
\includegraphics[height=4.1cm]{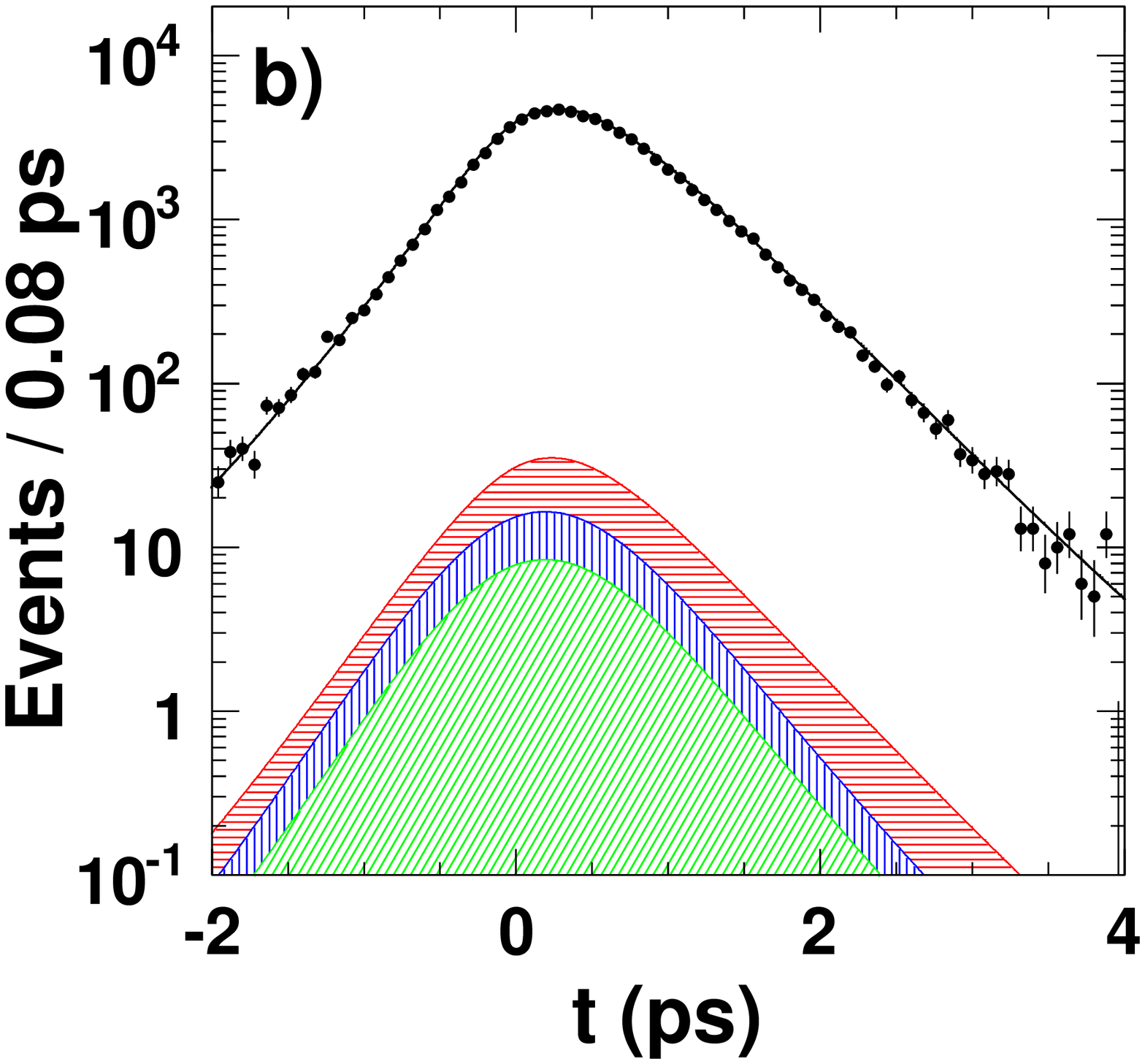} \\ 
\end{tabular}    
\caption{(color online) Proper-time distributions for (a) \kspipi and (b) \kskk  
data 
in the signal region with $-2<t<4$ \ps  
(points). 
The curves show the fit projections for signal plus background (solid lines) and for different background components  
(shaded regions). 
} 
\label{fig:properTime} 
\end{center} 
\end{figure}

Results for our nominal mixing fit, in which \Dz and \Dzb samples from 
\kspipi and \kskk channels are combined, are reported in Table~\ref{tab:results}. 
The proper-time distributions with their fit projections are shown in Fig.~\ref{fig:properTime}. 
Additional fit results and projections can be found in~\cite{ref:epaps}. 
We evaluate the amplitude model fit to the DP distribution with a $\chi^2$  
test with two-dimensional adaptive binning,  
and obtain $\chi^2= 10\hspace{2pt}429.2$ ($1\hspace{2pt}511.2$) for $8\hspace{2pt}626-41$ ($1\hspace{2pt}195-17$)  
degrees of freedom (ndof), for \kspipi (\kskk). 
MC studies show that a significant contribution to these $\chi^2$ values,  
$\Delta \chi^2/{\rm ndof} \approx 0.16$, arises from imperfections in modeling 
experimental effects, mostly the efficiency variations at the boundaries of the DP and the invariant mass resolution. 
The fitted average lifetime $\tau = 1/\Gamma$ is found to be consistent  
with the world average lifetime~\cite{ref:pdg2008}, while 
\tzero  
is found to be $5.1\pm0.8$ \fs ($5.1\pm2.2$ \fs) for  
the \kspipi (\kskk) mode, 
consistent with expectations  
from small misalignments in the detector~\cite{ref:Aubert:2009ck}. 
\begin{table}[hbt!] 
\caption{\label{tab:results} Results from the mixing fits. 
The first uncertainty is statistical, the second systematic and the third  
systematic from the amplitude model. For the nominal fit, the corresponding correlation coefficients 
between \x and \y are $3.5\%$, $16.0\%$ and $-2.7\%$, respectively. 
} 
\begin{ruledtabular} 
\begin{tabular}{lcc} 
\\[-0.15in] 
Fit type & $x/10^{-3}$ & $y/10^{-3}$ \\ 
\hline  
Nominal  & $\phantom{-1}1.6 \pm 2.3 \pm 1.2 \pm 0.8$ & $5.7 \pm 2.0 \pm 1.3 \pm 0.7$ \\ 
\kspipi  & $\phantom{-1}2.6 \pm 2.4$    & $6.0 \pm 2.1$ \\ 
\kskk    & $-13.6 \pm 9.2$              & $4.4 \pm 5.7$ \\ 
\Dz      & $\phantom{-1}0.0 \pm 3.3$    & $5.5 \pm 2.8$ \\ 
\Dzb     & $\phantom{-1}3.3 \pm 3.3$    & $5.9 \pm 2.8$  
\end{tabular} 
\end{ruledtabular} 
\end{table}

A variety of studies using large MC samples with both parameterized and full detector 
simulations and data have been performed to validate the analysis method  
and fitting procedure and to check the consistency of the results. 
These studies demonstrate that the analysis correctly determines the mixing 
parameters with insignificant biases and well-behaved Gaussian errors.  
No significant variations of the mixing parameters are observed 
 as a function of momentum, polar and azimuthal angles of the \Dz meson,  
and data taking period. 
Including the \mdz, \deltam PDFs in the mixing fit does not  
significantly change the values for \x and \y. 
The mixing fit has also been performed separately for the \kspipi and \kskk  
data samples, and for \Dz and \Dzb decays, with the results listed in Table~\ref{tab:results}. 
Fitting separately for \Dz and \Dzb provides a check against possible effects  
from \CPV in mixing and in decay.  
Finally, if we fit the data  
forcing the decay amplitudes for \Dz and \Dzb to be the same (no direct \CPV), but allowing their \x and \y values to differ,  
we find these values to be consistent (i.e., no evidence for \CPV in mixing).

\indent 
Systematic uncertainties arise from approximations in the modeling of experimental and selection criteria effects~\cite{ref:epaps}. 
We account for  
variations  
in the signal and background yields, 
the efficiency variations across the DP, 
the modeling of the DP and proper-time distributions for events containing misreconstructed \Dz decays, 
the misidentification of the \Dz flavor for signal and random \pipsoft events, 
potential effects  
due to mixing in the random \pipsoft background component, 
and PDF normalization.  
We also consider variations of the resolution function and \sigmat PDFs,  
including alternatives to describe the correlation between \sigmat and the DP position 
(e.g. neglecting the dependence of the \sigmat distributions on the DP position entirely). 
The dominant sources of experimental systematic uncertainty are  
the limited statistics of full detector simulations  
(used to study potential biases due to the event  
selection, invariant mass resolution, residual correlations between the fit variables, and fitting procedure) 
and instrumental effects arising from the small misalignment of the detector. 
Effects from our selection criteria are estimated by varying the \mdz, \deltam, \t, and \sigmat requirements.

Assumptions in the amplitude models are also a source of systematic uncertainty~\cite{ref:babar_gamma_dalitz2008,ref:epaps}. 
We use alternative models where the BW parameters are varied according to their uncertainties or changed 
to values measured by other experiments, the reference K-matrix solution~\cite{ref:babar_gamma_dalitz2008} is replaced by other solutions~\cite{ref:AS},  
and the standard parameterizations are substituted by other related choices. 
These include replacing the GS by BW lineshapes, removing the mass dependence in the P-vector~\cite{ref:Kmatrix-Pvector}, 
changes in form factors such as  
variations 
in the Blatt-Weisskopf radius 
and the effect of evaluating the momentum of the spectator particle in the \Dz meson frame rather than in the resonance rest frame, 
and adopting a helicity formalism~\cite{ref:RevDalitzPlotFormalism} to describe the angular dependence. 
Other models are built  
by removing or adding resonances with small or negligible fractions.  
The largest effect arises when the  
uncertainties in the amplitude model parameters  
obtained from the fit to the DP variables only~\cite{ref:epaps} are propagated to the mixing fit. 
These uncertainties are dominated by the parameters related to the $\K\pi$ S and P waves. 
%

\begin{figure}[htb] 
\begin{center} 
\includegraphics[width=0.45\textwidth]{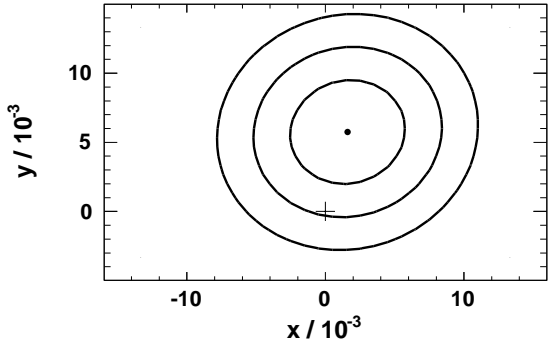} 
\caption{Central value (point) and C.L. contours  
(including statistical, systematic and amplitude model uncertainties)  
in the \x-\y plane for  
\mbox{C.L.} = 68.3\%, 95.4\%, 99.7\%. 
The no-mixing point is shown as a plus sign ($+$). 
} 
\label{fig:contours} 
\end{center} 
\end{figure}

\indent 
The mixing significance is evaluated by the variation of the negative log-likelihood  
(\twoDLL)  
in the mixing parameter space. 
We account for the systematic uncertainties  
by approximating \Lik as a two-dimensional Gaussian with covariance matrix resulting 
from the sum of the corresponding statistical, systematic, and amplitude model matrices. 
Figure~\ref{fig:contours} shows the confidence-level (C.L.) contours in two dimensions (\x and \y)  
with systematic uncertainties included. 
The variation in \twoDLL for the no-mixing point is  
$5.6$ units 
which  
corresponds to a 
C.L. 
equivalent to $1.9$  
standard deviations,  
including the systematic uncertainties.     

\indent 
In summary,  
we have directly  
measured the mixing parameters  
$x= [1.6 \pm 2.3 \hbox{ (stat.)} \pm 1.2 \hbox{ (syst.)} \pm 0.8 \hbox{ (model)} ] \times10^{-3}$,  
and   
$y= [5.7 \pm 2.0 \hbox{ (stat.)} \pm 1.3 \hbox{ (syst.)} \pm 0.7 \hbox{ (model)} ]\times 10^{-3}$, 
using, for the first time,  
a combined analysis of $\Dztokspipi$ and $\Dztokskk$ decays. 
These results are consistent with the  
previous similar measurements of \kspipi alone~\cite{ref:Asner:2005sz,ref:Abe:2007rd} and have improved precision. 
They disfavor 
the no-mixing hypothesis with a  
C.L. 
equivalent to $1.9$ standard deviations and 
are in agreement with the range of SM predictions~\cite{ref:Petrov:2006nc,ref:Falk:2001hx,ref:Bigi:2000wn,ref:Wolfenstein:1985ft,ref:Falk:2004wg}. 
Our  
measurements favor 
lower values for \x than for \y,  
and lower \x and \y values (but still consistent) than those obtained  
when combining results from  
other \Dz decays~\cite{ref:Aubert:2007wf,ref:Staric:2007dt,ref:Aaltonen:2007uc,ref:Aubert:2007en,ref:Aubert:2009ck,ref:Aubert:2008zh,ref:CLEOc-KpiPhase}. 
Adding our results to the combination of all previous analyses significantly improves our 
current knowledge of the mixing parameters \x and \y, 
whose average values  
change from $(9.8\pm2.5)\times 10^{-3}$ and $(8.3\pm1.6)\times 10^{-3}$  
to $(5.9\pm2.0)\times 10^{-3}$ and $(8.0\pm1.3)\times 10^{-3}$, respectively~\cite{ref:hfagFPCP10}.

We are grateful for the excellent luminosity and machine conditions
provided by our \pep2\ colleagues, 
and for the substantial dedicated effort from
the computing organizations that support \babar.
The collaborating institutions wish to thank 
SLAC for its support and kind hospitality. 
This work is supported by
DOE
and NSF (USA),
NSERC (Canada),
CEA and
CNRS-IN2P3
(France),
BMBF and DFG
(Germany),
INFN (Italy),
FOM (The Netherlands),
NFR (Norway),
MES (Russia),
MICIIN (Spain),
STFC (United Kingdom). 
Individuals have received support from the
Marie Curie EIF (European Union),
the A.~P.~Sloan Foundation (USA)
and the Binational Science Foundation (USA-Israel).

\onecolumngrid 
\newpage 
 
\setcounter{page}{1} 
\setcounter{table}{0} 
\setcounter{figure}{0} 
 
\begin{center} 
{\large \bfseries \boldmath 
Measurement of \Dz-\Dzb mixing parameters using $\Dz \to \KS\pip\pim$ and $\Dz \to \KS\Kp\Km$ decays}\\ 
The \babar\ Collaboration 
\end{center}

\begin{center} 
The following includes supplementary material for the Electronic 
Physics Auxiliary Publication Service.  
\end{center}

\begin{table}[h!] 
\caption{\label{tab:kspipimodels} 
\Dztokspipi complex amplitudes,  
$\pi\pi$ P-vector and $K\pi$ S-wave parameters,  
and  
fit fractions, 
as obtained from the  
mixing fit. 
The $\pi\pi$ S-wave parameters $\beta_5$, $f_{14}^{\rm prod}$, and $f_{15}^{\rm prod}$ are fixed to zero due to the lack of sensitivity.  
We also report the mass and the width of the $K^*(892)^\mp$ resonance. 
Errors are statistical only.  
The fit fraction is defined as the integral over the entire DP of a single component  
divided by the coherent sum of all components.  
The sum of fit fractions is $103.3\%$. 
A detailed description of the parameters can be found elsewhere~\cite{ref:babar_gamma_dalitz2008}. 
Equations~(14) and (15) in~\cite{ref:babar_gamma_dalitz2008} have been corrected as follows,  
${\cal A}_{\K\pi\ L=0}(\s) = T_{\K\pi\ L=0}(\s)/\rho(s)$,  
where 
$\rho(s) = q/\sqrt{s}$ is the phase-space factor and 
$T_{\K\pi\ L=0}(\s) = F \sin(\delta_F + \phi_F) e^{i(\delta_F+\phi_F)} + R \sin\delta_R e^{i(\delta_R+\phi_R)}e^{i2(\delta_F+\phi_F)}$, 
with  
$\tan \delta_R = M_{K^{*}_0(1430)} \Gamma_{K^{*}_0(1430)}(s)/(M_{K^{*}_0(1430)}^2-s)$, 
$\cot \delta_F = 1/(aq)+rq/2$, 
\s the invariant mass squared of the $\K\pi$ system,  
and $q$ the momentum of the kaon (or pion) in the $\K\pi$ rest frame~\cite{ref:LASS}. 
The symbol $^\dagger$ indicates the parameters fixed in the mixing fit to the values extracted  
from a time-integrated DP fit to the same data. 
The results from this time-integrated DP fit for the amplitude model parameters agree within statistical errors with the results reported here.  
} 
\begin{ruledtabular} 
\begin{tabular}{lccc} 
\\[-0.15in] 
    Component  &  Amplitude   &  Phase (rad) & Fit fraction (\%) \\ [0.01in] 
\hline 
$K^{*}(892)^-$        &    $1.735\pm0.005\phdag$                  & $\phm2.331\pm0.004\phdag$             & $57.0$ \\   
$\rho(770)^0$         &    $1\phdag$                              & $\phm0\phdag$                         & $21.1$ \\  
 
$K^{*}_0(1430)^-$     &    $\phz\phz2.650\pm0.015\phz\phz\phdag$  & $\phm1.497\pm0.007\phdag$                & $\phz6.1$ \\   
$K^{*}_2(1430)^-$     &    $1.303\pm0.013\phdag$                  & $\phm2.498\pm0.012\phdag$                  & $\phz1.9$ \\   
$\omega(782)$         &    $0.0420\pm0.0006\phdag$                & $\phm2.046\pm0.014\phdag$                & $\phz0.6$ \\ 
$K^{*}(892)^+$        &    $0.164\pm0.003\phdag$                  & $-0.768\pm0.019\phdag$                  & $\phz0.6$ \\ 
$K^{*}(1680)^-$       &    $\phz0.90\pm0.03\phz\phdag$            & $-2.97\pm0.04\phdag$                  & $\phz0.3$ \\ 
$f_2(1270)$           &    $0.410\pm0.013\phdag$                  & $\phm2.88\pm0.03\phdag$                   & $\phz0.3$ \\ 
$K^{*}_0(1430)^+$     &    $\phz0.145\pm0.014\phz\phdag$          & $\phm1.78\pm0.10\phdag$                    & $<0.1\phz$ \\ 
$K^{*}_2(1430)^+$     &    $0.115\pm0.013\phdag$                  & $\phm2.69\pm0.11\phdag$                    & $<0.1\phz$ \\ 
\\ 
$\lceil$ 
$\pi\pi$ S-wave     &                 &            &  $15.4$ \\ 
$\phm\beta_1$             &    $5.54\pm0.06\phdag$               & $-0.054\pm0.007\phdag$   &  \\ 
$\phm\beta_2$             &    $15.64\pm0.06\phz\phdag$          & $-3.125\pm0.005\phdag$ &   \\  
$\phm\beta_3$             &    $44.6\pm1.2\phz\phdag$            & $\phm2.731\pm0.015\phdag$ & \\  
$\phm\beta_4$             &    $9.3\pm0.2\phdag$                 & $\phm2.30\pm0.02\phdag$ &\\  
$\phm f_{11}^{\rm prod}$  &    $11.43\pm0.11^\dagger\phz$        & $-0.005\pm0.009^\dagger$  &\\  
$\phm f_{12}^{\rm prod}$  &    $15.5\pm0.4^\dagger\phz$          & $-1.13\pm0.02^\dagger$   & \\ 
$\phm f_{13}^{\rm prod}$       &    $\phz 7.0\pm0.7^\dagger\phz$ & $\phm0.99\pm0.11^\dagger$    & \\ 
                          & \multicolumn{2}{c}{Parameter value} & \\ 
$\lfloor$  
$s_0^{\rm prod}$          & \multicolumn{2}{c}{$-3.92637$} & \\ 
\\ 
$\lceil$ 
$K\pi$ S-wave parameters               & \multicolumn{2}{c}{}         & \\ 
$\phm M_{K^{*}_0(1430)}$~(\mevcc)          &  \multicolumn{2}{c}{$1421.5\pm1.6^\dagger$}      & \\ 
$\phm \Gamma_{K^{*}_0(1430)}$~(\mevcc)     &  \multicolumn{2}{c}{$\phz247\pm3^\dagger$}     &  \\ 
$\phm F$                   &  \multicolumn{2}{c}{$\phz\phz\phz0.62\pm0.04^\dagger$}     &\\  
$\phm \phi_F$ (rad)        &  \multicolumn{2}{c}{$\phz\phz-0.100\pm0.010^\dagger\phm$}     &\\  
$\phm R$                   &  \multicolumn{2}{c}{$\phz\phz\phz1\phdag$}       &\\  
$\phm \phi_R$ (rad)        &  \multicolumn{2}{c}{$\phz\phz\phz1.10\pm0.02^\dagger$}    &\\  
$\phm a$ (\invgevc)        &  \multicolumn{2}{c}{$\phz\phz\phz0.224\pm0.003^\dagger$}     &\\  
$\lfloor$ 
$r$ (\invgevc)             &  \multicolumn{2}{c}{$\phz-15.01\pm0.13^\dagger\phm$}        &\\  
\\ 
$\lceil$ 
$K^{*}(892)$ parameters                &  \multicolumn{2}{c}{}        &  \\ 
$\phm M_{K^{*}(892)}$~(\mevcc)          &  \multicolumn{2}{c}{$\phm893.70\pm0.07^\dagger$}    &  \\ 
$\lfloor$ 
$\Gamma_{K^{*}(892)}$~(\mevcc)     &  \multicolumn{2}{c}{$\phm\phz46.74\pm0.15^\dagger$}     &  \\ 
\end{tabular} 
\end{ruledtabular} 
\end{table}

\begin{table}[h!] 
\caption{\label{tab:kskkmodel} 
\Dztokskk complex amplitudes 
and  
fit 
fractions, 
as obtained from the  
mixing fit. 
We also report the mass and the width of the $\phi(1020)$ resonance,  
and the $a_0(980)$ coupling constant to $\K\Kb$ as determined from the fit. 
Errors are statistical only.  
The fit fraction is defined as the integral over the entire DP of a single component  
divided by the coherent sum of all components.  
The sum of fit fractions is $163.4\%$. 
A detailed description of the parameters can be found elsewhere~\cite{ref:babar_gamma_dalitz2008}. 
The symbol $^\dagger$ indicates the parameters fixed in the mixing fit to the values extracted  
from a time-integrated DP fit to the same data. 
The results from this time-integrated DP fit for the amplitude model parameters agree within statistical errors with the results reported here.  
} 
\begin{ruledtabular} 
\begin{tabular}{lccc} 
\\[-0.15in] 
    Component  &  Amplitude   &  Phase (rad) & Fit fraction (\%) \\ [0.01in] 
\hline 
$a_0(980)^0$      &  $1\phdag$                        &  $\phm0\phdag$                       & $51.8$  \\ 
$\phi(1020)$      &  $0.2313\pm0.0011\phdag$          &  $-0.977\pm0.008\phdag$          & $44.1$   \\ 
$a_0(1450)^+$     &  $\phz0.93\pm0.03^\dagger\phz$    &  $\phm1.66\pm0.07^\dagger$           & $25.6$  \\ 
$a_0(980)^+$      &  $0.635\pm0.006\phdag$            &  $-2.91\pm0.02\phdag$                & $19.5$  \\ 
$a_0(1450)^0$     &  $\phz0.83\pm0.10^\dagger\phz$    &  $-1.93\pm0.12^\dagger$              & $19.3$       \\ 
$f_0(1370)$       &  $0.16\pm0.05^\dagger$            &  $\phm0.2\pm0.2^\dagger$             & $\phz1.7$  \\ 
$f_2(1270)$       &  $\phz0.385\pm0.015\phz\phdag$    &  $\phm0.06\pm0.04\phdag$                & $\phz0.7$      \\ 
$a_0(980)^-$      &  $0.125\pm0.008\phdag$            &  $\phm2.47\pm0.04\phdag$             & $\phz0.7$       \\ 
\\ 
$\lceil$ 
$\phi(1020)$ and $a_0(980)$ parameters   & \multicolumn{2}{c}{Value}               & \\ 
$\phm M_{\phi(1020)}$~(\mevcc)          &  \multicolumn{2}{c}{$1019.55\pm0.02^\dagger$}           &  \\ 
$\phm \Gamma_{\phi(1020)}$~(\mevcc)         &  \multicolumn{2}{c}{$\phz\phz\phz4.60\pm0.04^\dagger$}   &  \\ 
$\lfloor$ 
$g_{\K\Kb}$~(\mevcc)                  &  \multicolumn{2}{c}{$\phz537\pm9^\dagger$}         & \\ 
\end{tabular} 
\end{ruledtabular} 
\end{table}


\begin{figure}[!h] 
\begin{center} 
\begin{tabular} {cc}   
\includegraphics[height=5.1cm]{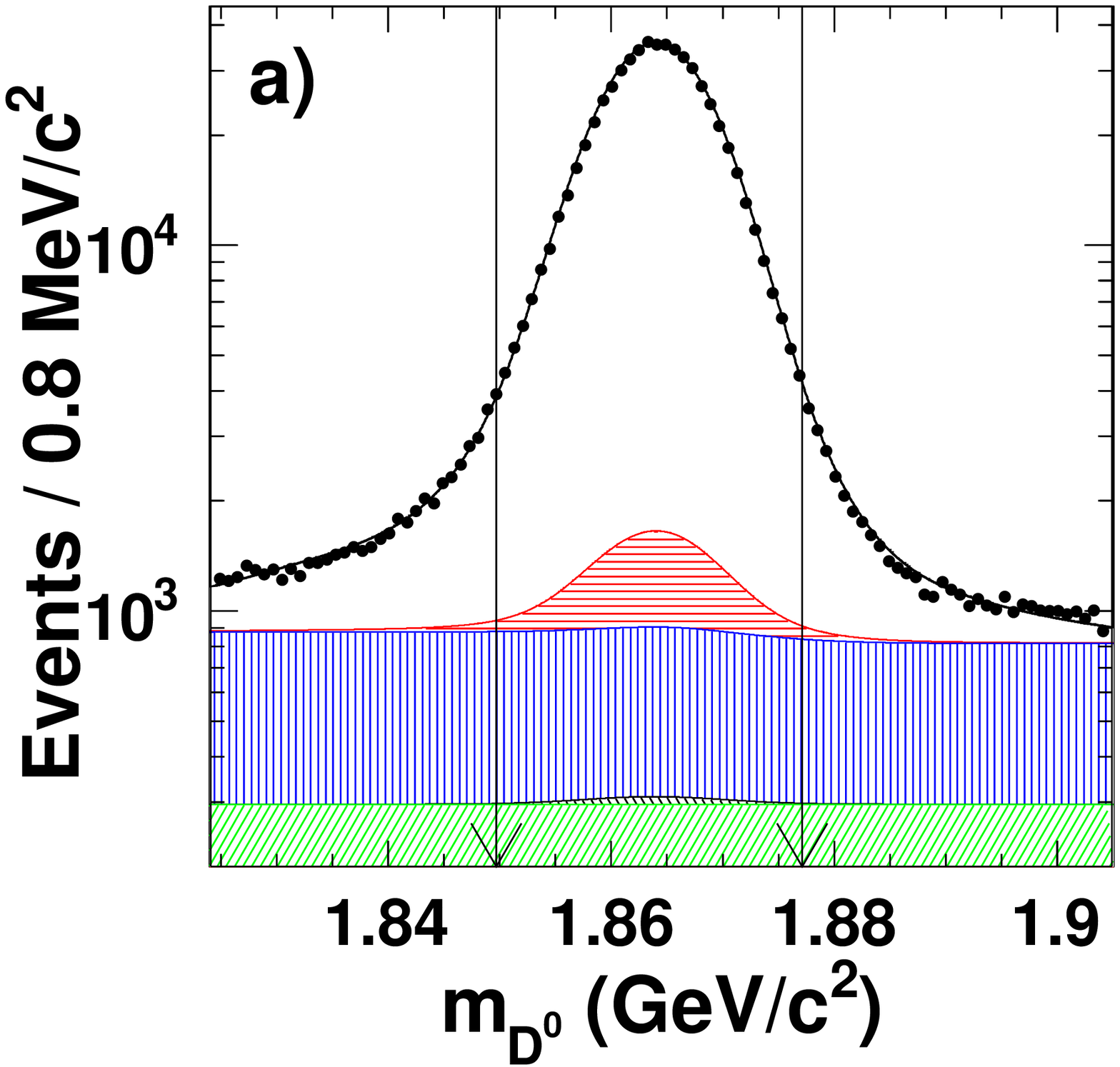} & 
\includegraphics[height=5.1cm]{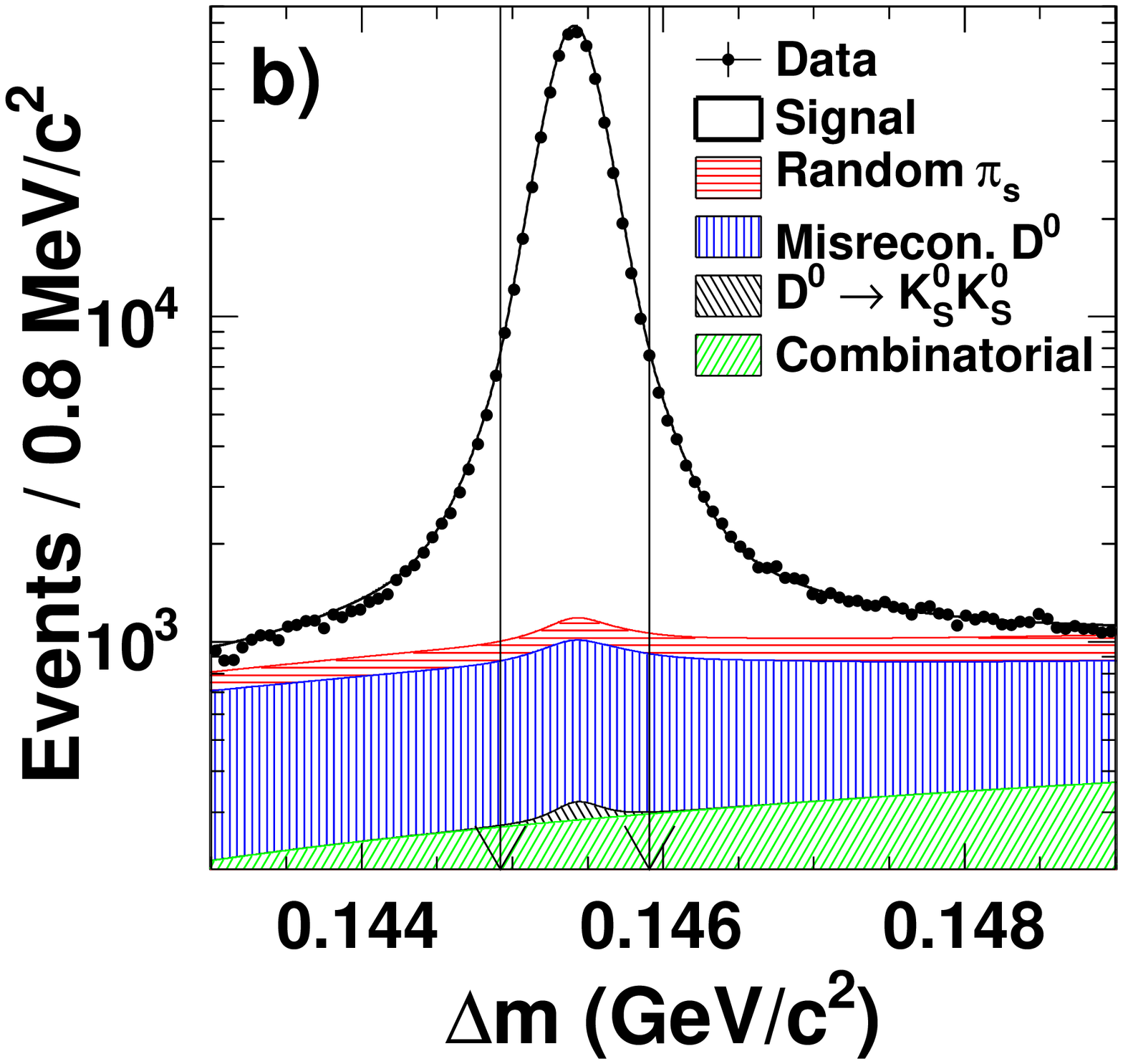} \\ 
\includegraphics[height=5.1cm]{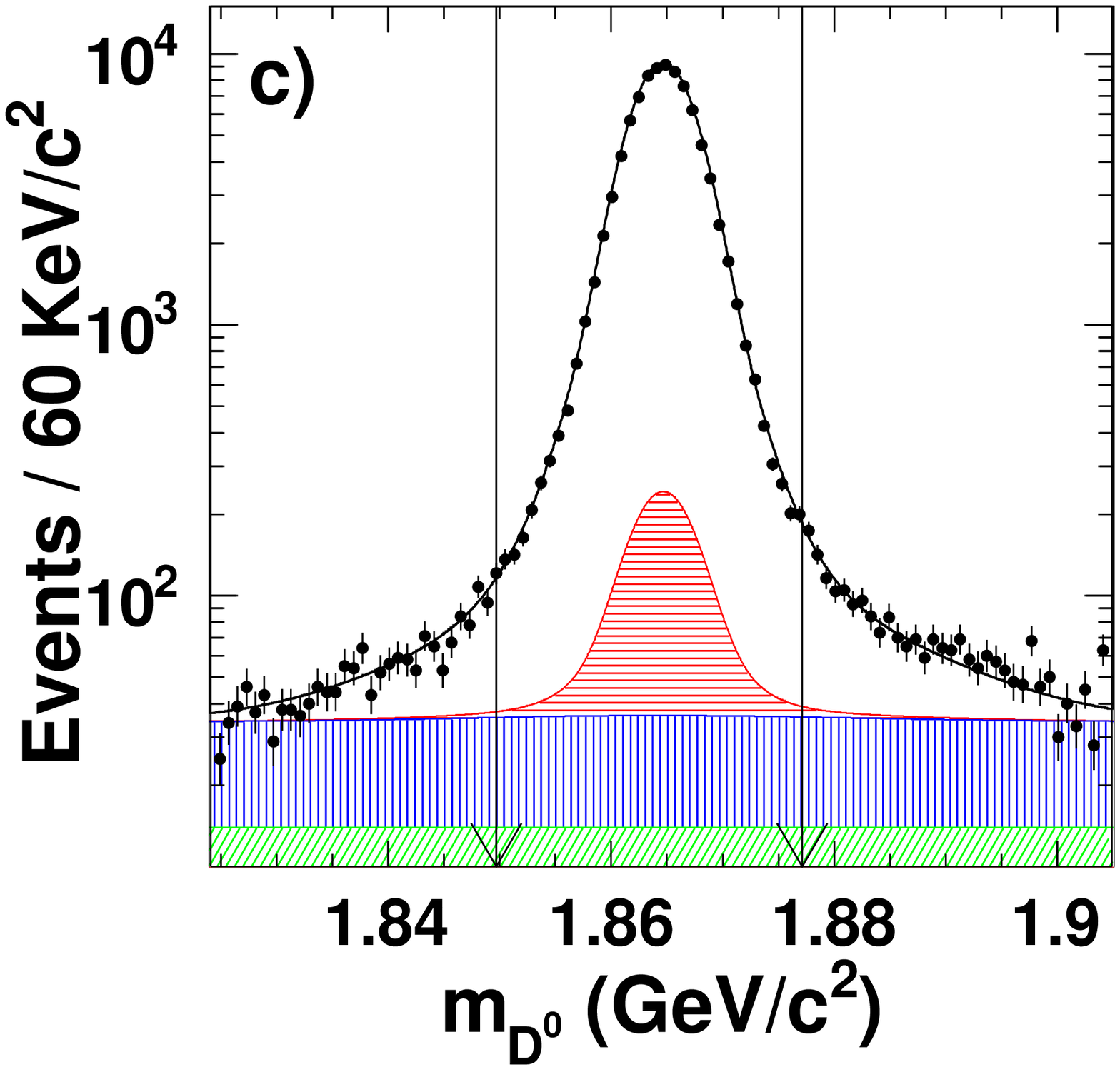} & 
\includegraphics[height=5.1cm]{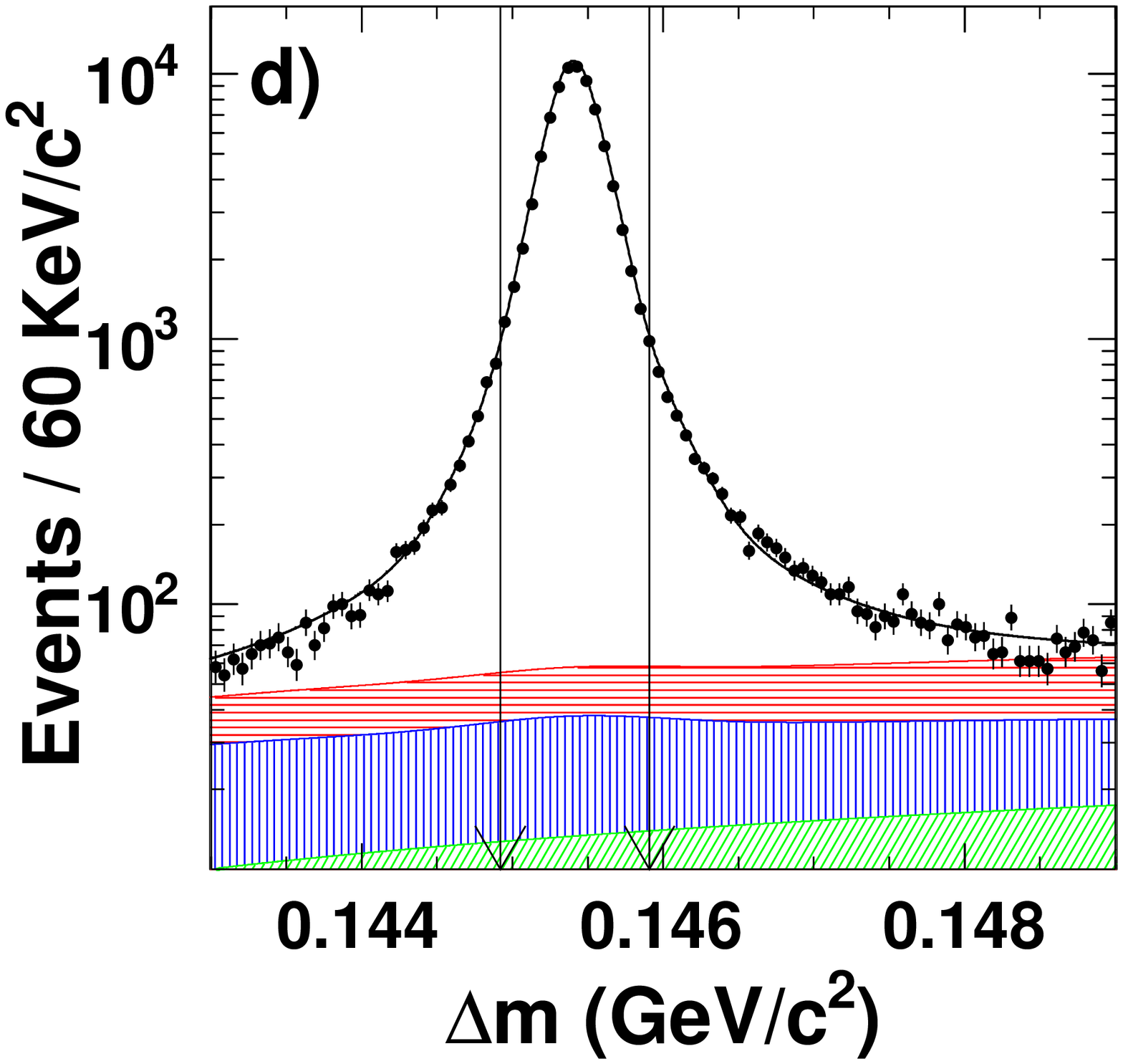} \\ 
\end{tabular}    
\caption{(color online) Distributions of \mdz  and \deltam for  
(a,b) \kspipi 
and (c,d) \kskk data 
after all selection criteria  
(points).  
The curves superimposed represent the fit projections  
for signal plus background (solid lines) and for different background components (shaded regions). 
The arrows indicate the definition of the signal region. 
} 
\label{fig:mD-DeltaM-projections} 
\end{center} 
\end{figure}

\begin{figure}[h!] 
\begin{tabular} {cc} 
{\includegraphics[height=5.7cm]{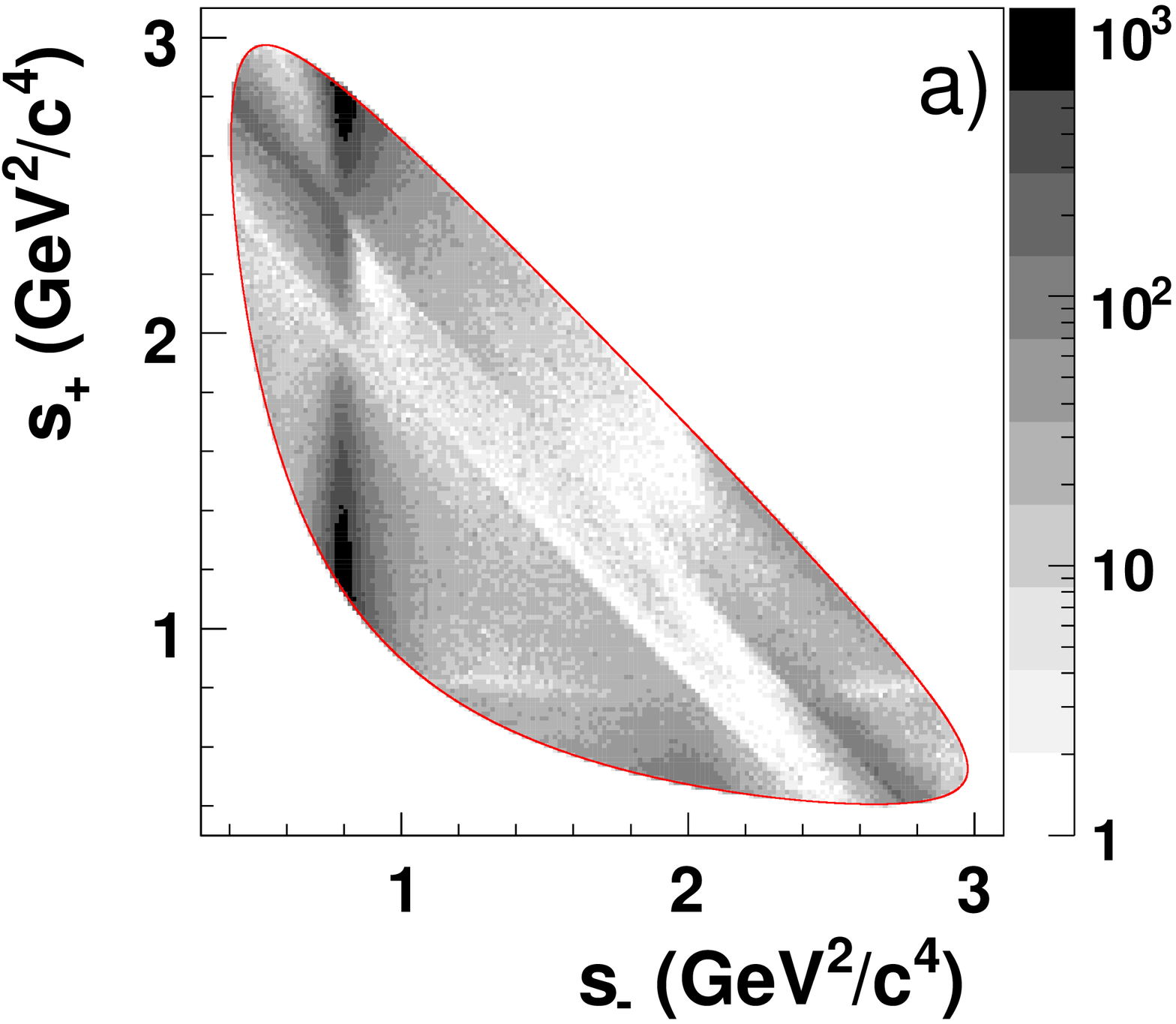}} & 
{\includegraphics[height=5.7cm]{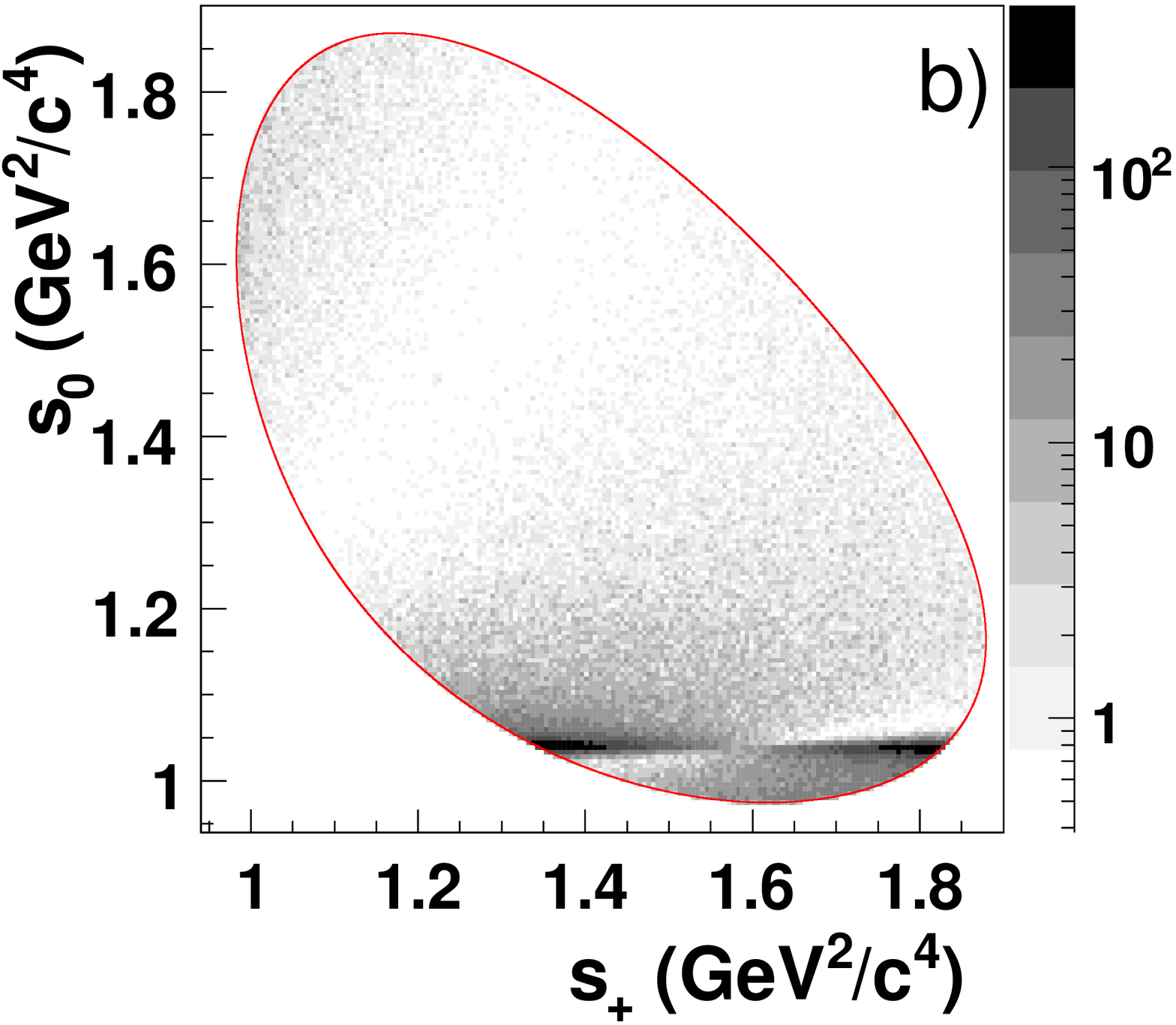}} \\ 
\end{tabular}    
\caption{\label{fig:Dalitz} DP distributions for (a) \Dztokspipi and (b) \Dztokskk  
data 
after all selection criteria, in the signal region.  
The gray scale indicates the number of events per bin.  
The solid lines show the kinematic limits of the \Dz decay.  
The \szero DP variable is defined as $\szero = m^2(h^+h^-)$. 
For \Dzb decays the variables \sminus and \splus are interchanged.  
} 
\end{figure}

\begin{figure*}[h!] 
\begin{center} 
\begin{tabular} {ccc}   
\includegraphics[height=3.8cm]{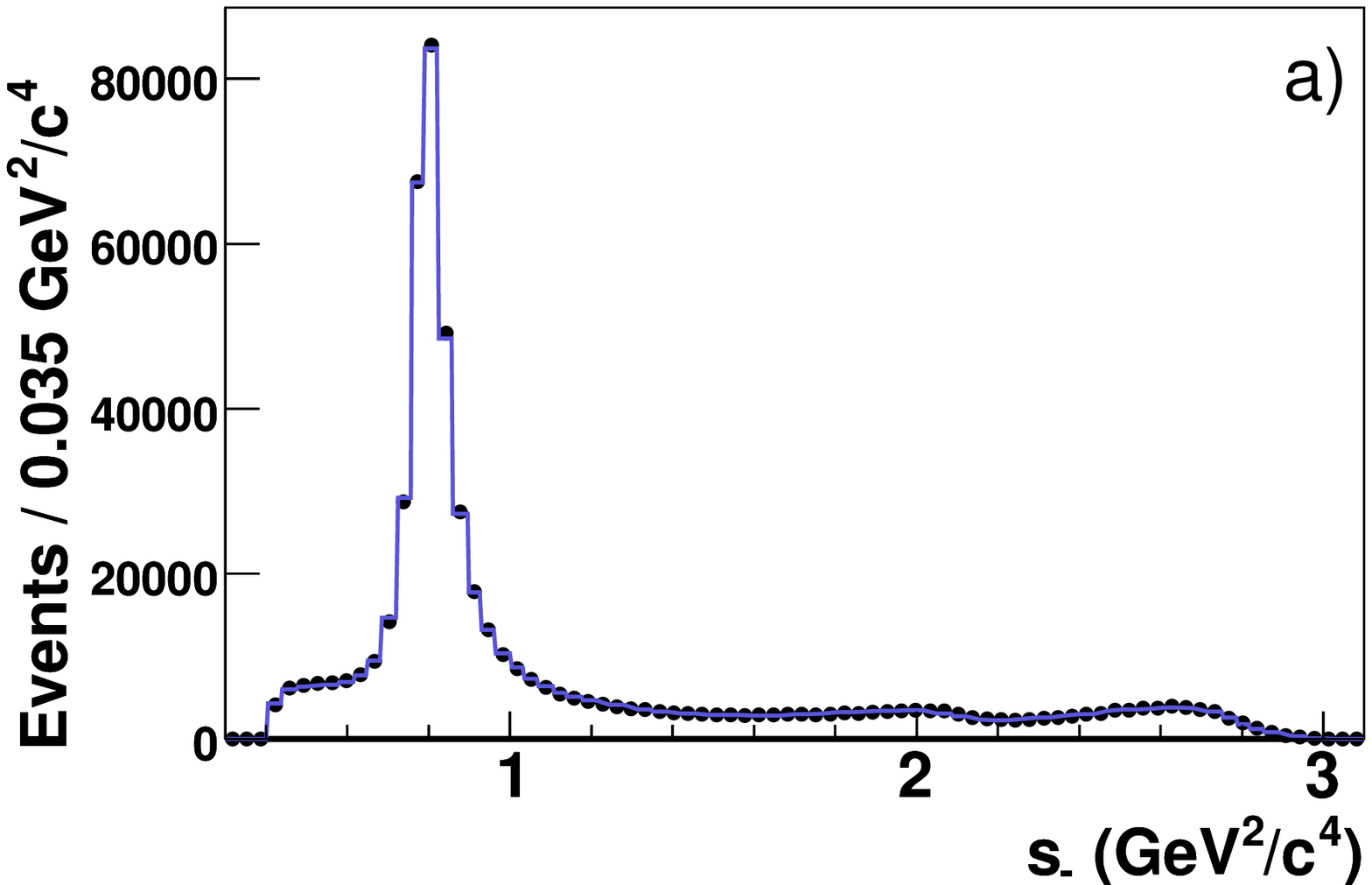} & 
\includegraphics[height=3.8cm]{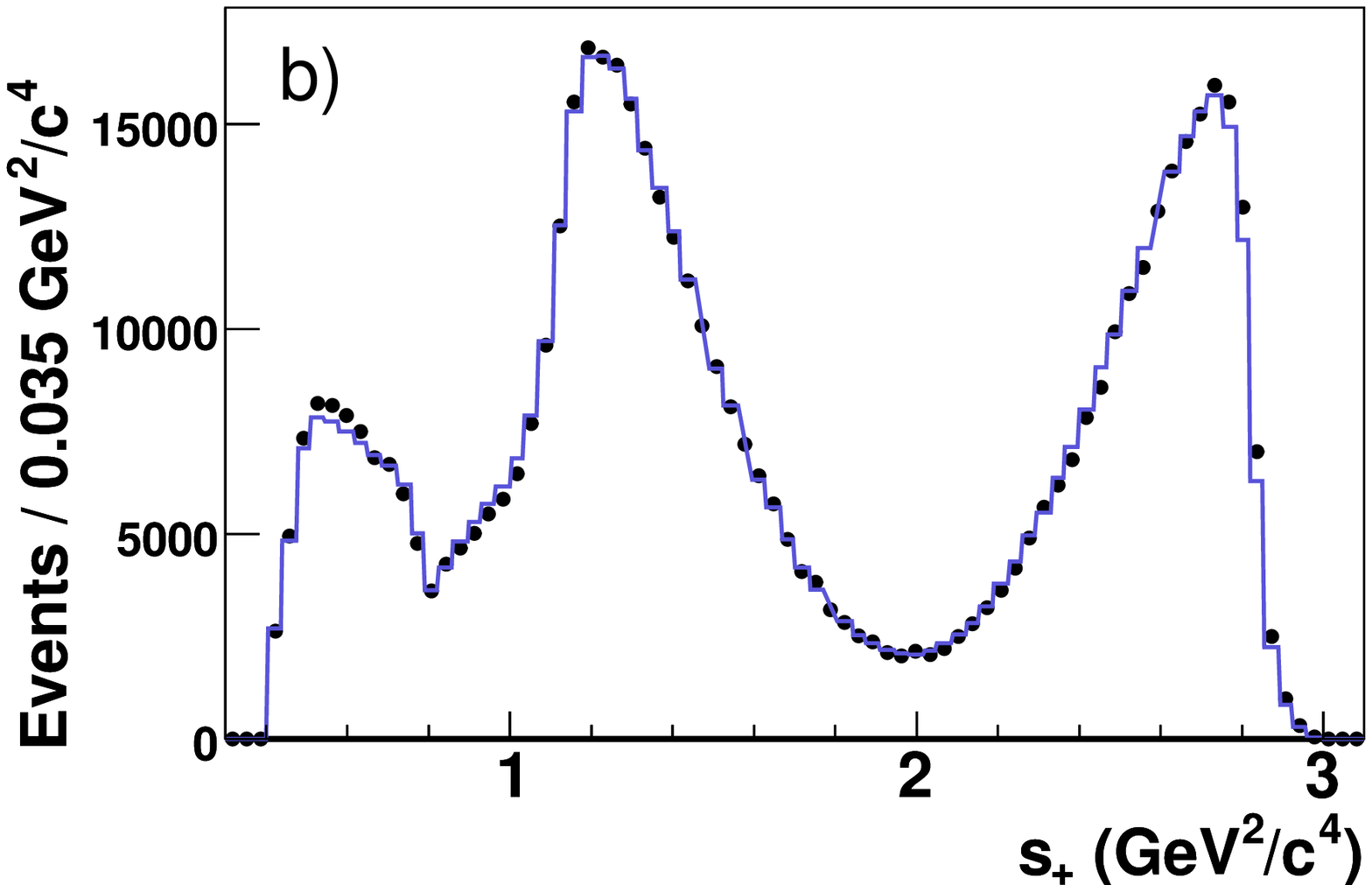} & 
\includegraphics[height=3.8cm]{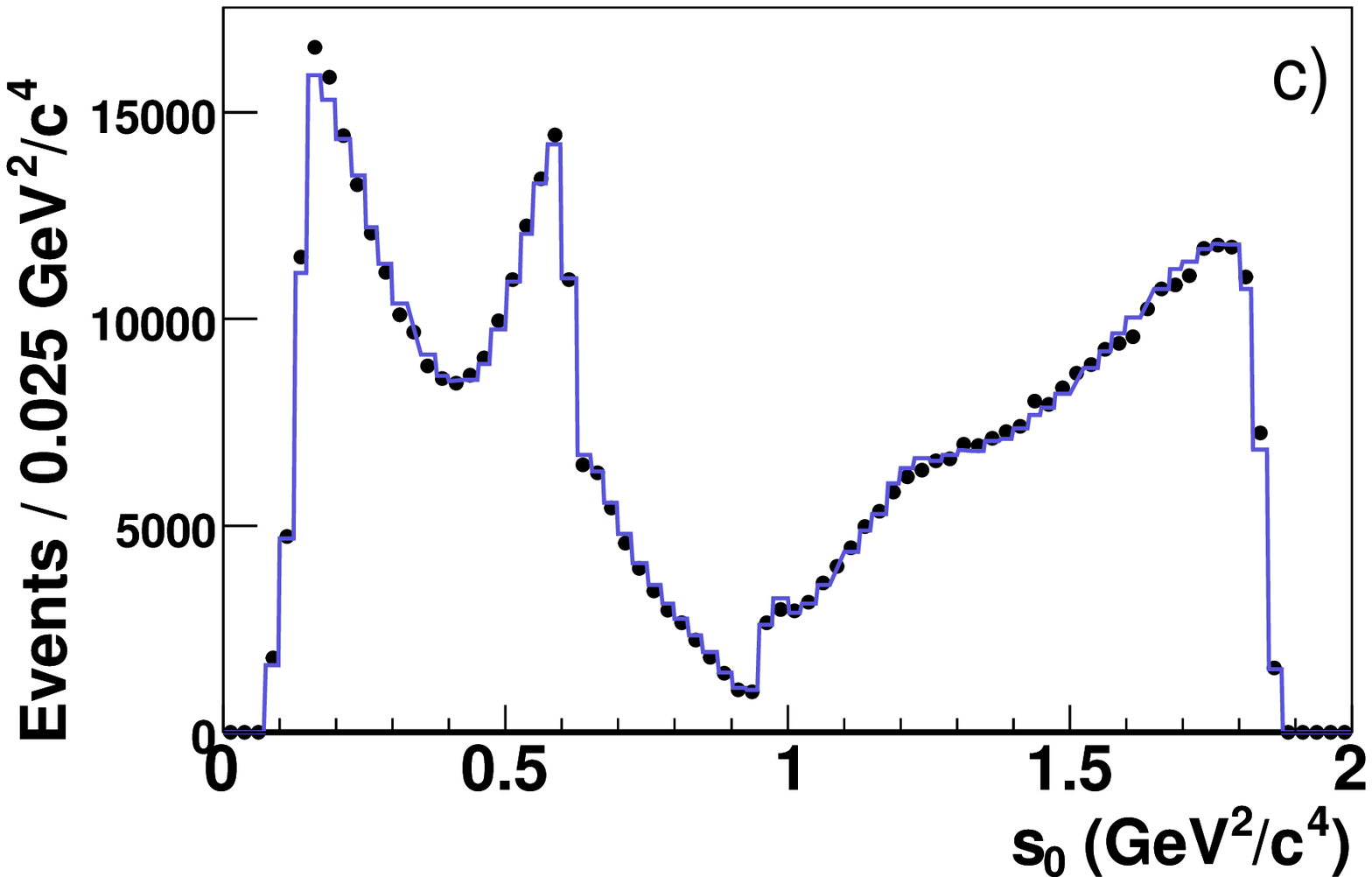} \\ 
\includegraphics[height=3.8cm]{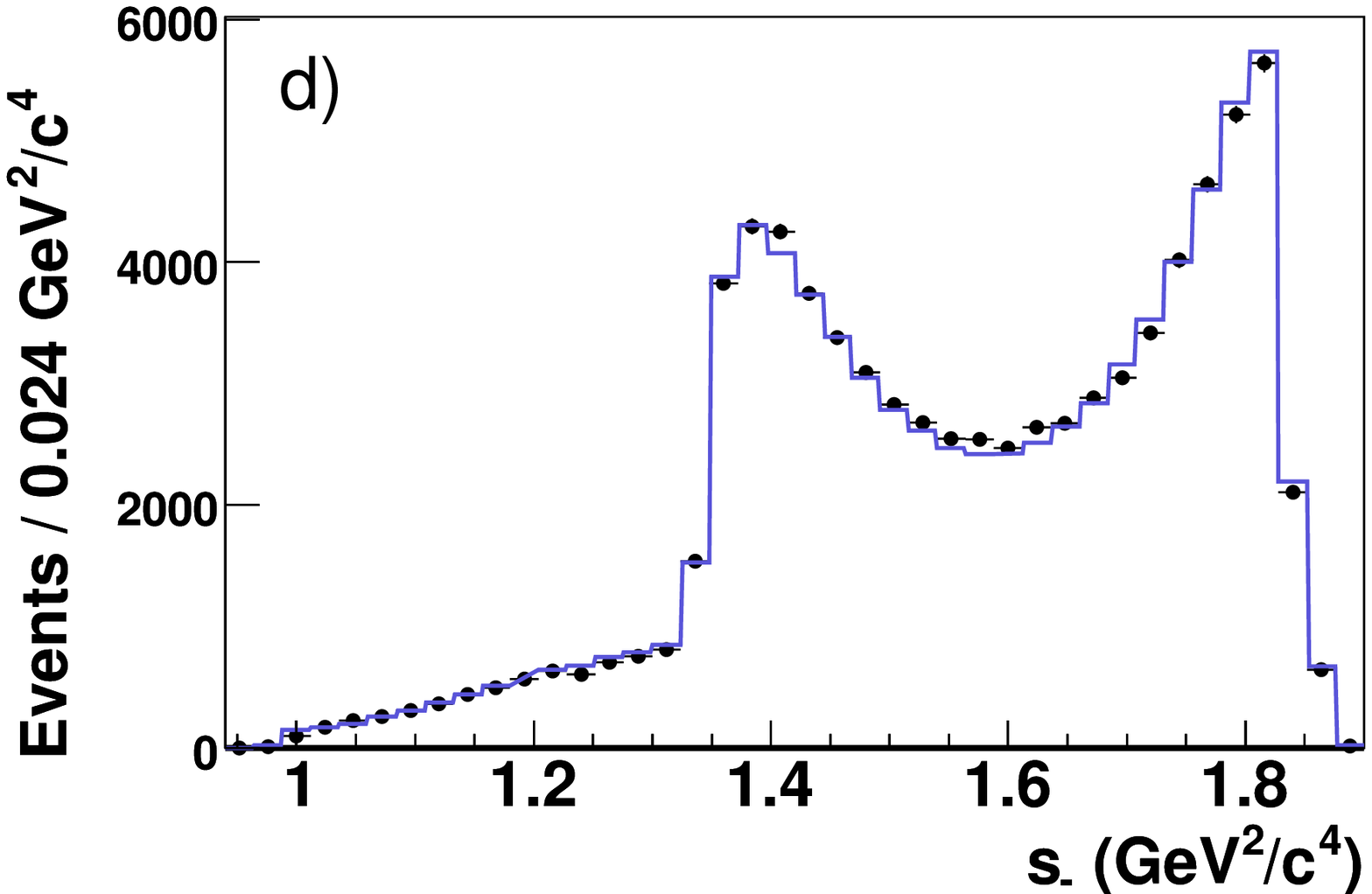} & 
\includegraphics[height=3.8cm]{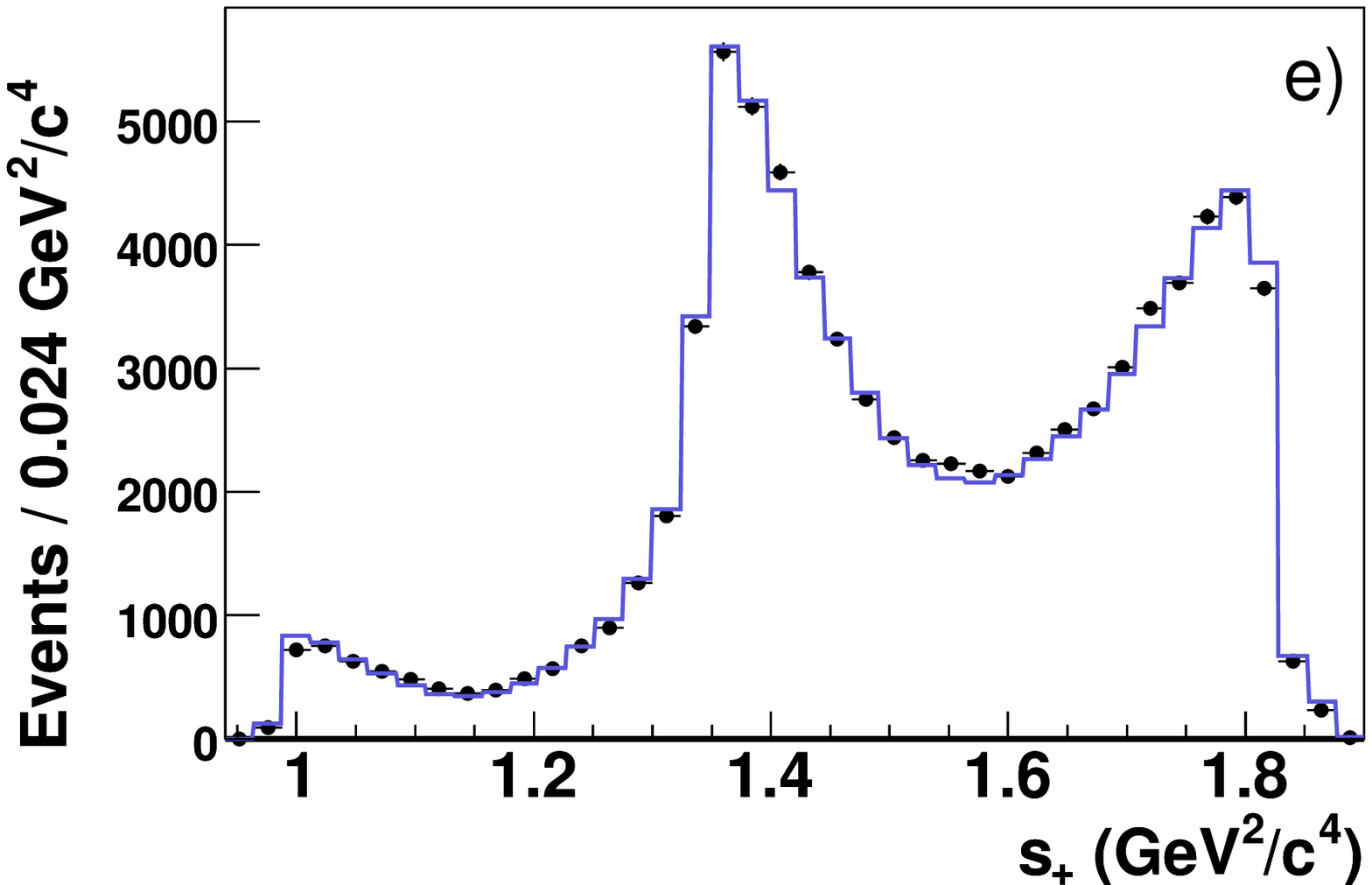} & 
\includegraphics[height=3.8cm]{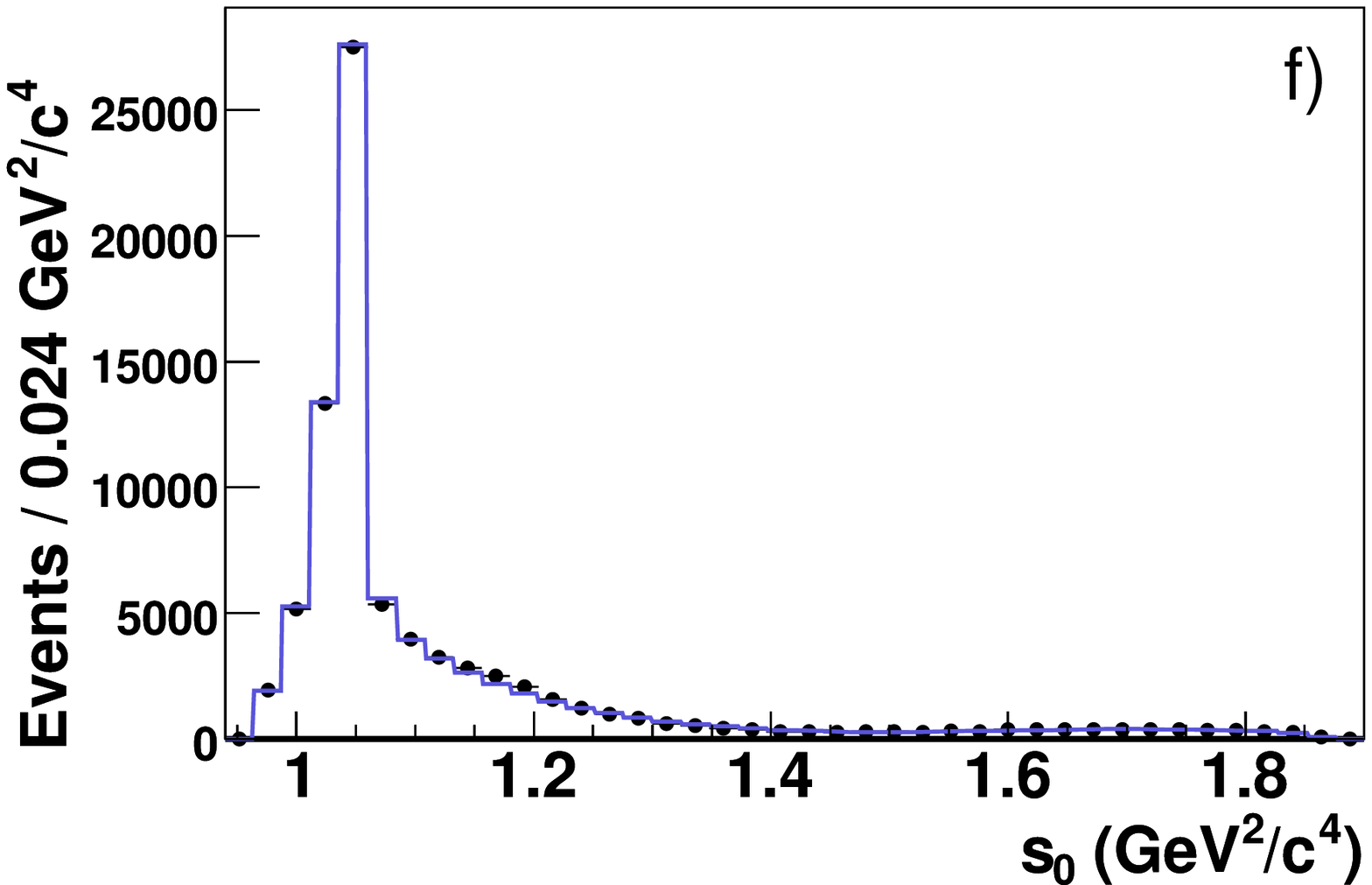} \\ 
\end{tabular}    
\caption{\label{fig:DalitzProjections} DP projections for (a,b,c) \Dztokspipi and (d,e,f) \Dztokskk  
data after all selection criteria, in the signal region (points).  
The histograms represent the mixing fit projections. 
For \Dzb decays the variables \sminus and \splus are interchanged. 
} 
\end{center} 
\end{figure*}

\onecolumngrid 
\newpage


\begin{table}[hbt!] 
\caption{\label{tab:cartesian-exp-syst}  
Summary of the contributions to the experimental systematic uncertainty on the mixing parameters. 
} 
\begin{ruledtabular} 
\begin{tabular}{lcc} 
\\[-0.15in] 
 Source                               & $\x/10^{-3}$  & $\y/10^{-3}$  \\ [0.01in] \hline 
 Analysis biases and fitting procedure (Monte Carlo statistics)                             & 0.75 & 0.66 \\ 
 Selection criteria                                                                         & 0.47 & 0.57 \\ 
 Signal and background yields                                                               & 0.11 & 0.07 \\ 
 Efficiency variations across the DP                                                        & 0.37 & 0.18 \\ 
 Modeling of the DP distributions for misreconstructed \Dz decays                           & 0.33 & 0.14 \\ 
 Modeling of the proper-time distributions for signal and misreconstructed \Dz decays       & 0.13 & 0.13 \\ 
 Modeling of the proper-time error distributions for signal and misreconstructed \Dz decays & 0.06 & 0.09 \\ 
 Misidentification of the \Dz flavor for signal and random \pipsoft events                  & 0.49 & 0.40 \\ 
 Mixing in the random \pipsoft background component                                         & 0.10 & 0.08 \\ 
 PDF normalization                                                                          & 0.11 & 0.05 \\ 
 Misalignment of the detector                                                               & 0.28 & 0.83 \\ 
 \hline  
 Total experimental systematic uncertainty                                                  & 1.18 & 1.30 \\ 
\end{tabular} 
\end{ruledtabular} 
\end{table}

\begin{table}[hbt!] 
\caption{\label{tab:cartesian-model-syst}  
Summary of the contributions to the \Dz decay amplitude model systematic uncertainty on the mixing parameters. 
} 
\begin{ruledtabular} 
\begin{tabular}{lcc} 
\\[-0.15in] 
 Source                               & $\x/10^{-3}$  & $\y/10^{-3}$  \\ [0.01in] \hline 
 Breit-Wigner parameters and alternative GS lineshapes          & 0.35 & 0.12 \\ 
 Alternative K-matrix solutions and P-vector parameterization   & 0.13 & 0.19 \\ 
 $\K\pi$ S- and P-waves, and $\pi\pi$ S-wave parameters         & 0.68 & 0.53 \\  
 Form factors                                                   & 0.25 & 0.23 \\ 
 Angular dependence                                             & 0.05 & 0.17 \\ 
 Add/remove resonances                                          & 0.17 & 0.23 \\ 
\hline 
 Total amplitude model systematic uncertainty           & 0.83 & 0.69 \\ 
\end{tabular} 
\end{ruledtabular} 
\end{table}


\begin{thebibliography}{99} 
 
 
 
\bibitem{ref:Petrov:2006nc} 
  A.~A.~Petrov, 
  Int.\ J.\ Mod.\ Phys.\  A {\bf 21}, 5686 (2006). 
 
 
\bibitem{ref:Falk:2001hx} 
  A.~F.~Falk, Y.~Grossman, Z.~Ligeti and A.~A.~Petrov, 
  \jprd{65}, 054034 (2002). 
 
\bibitem{ref:Bigi:2000wn} 
  I.~I.~Y.~Bigi and N.~G.~Uraltsev, 
  \npb{592},  92 (2001). 
 
\bibitem{ref:Wolfenstein:1985ft} 
  L.~Wolfenstein, 
  \plb{164}, 170 (1985). 
 
 
\bibitem{ref:Falk:2004wg} 
A.~F.~Falk, Y.~Grossman, Z.~Ligeti, Y.~Nir and A.~A.~Petrov, 
\jprd{69}, 114021 (2004). 
 
 
 
\bibitem{ref:Aubert:2007wf} 
  B.~Aubert {\it et al.} (\babar\ Collaboration), 
  \jprl{98}, 211802 (2007). 
 
\bibitem{ref:Staric:2007dt} 
  M.~Staric {\it et al.} (Belle Collaboration), 
  \jprl{98}, 211803 (2007). 
 
\bibitem{ref:Aaltonen:2007uc} 
  T.~Aaltonen {\it et al.} (CDF Collaboration), 
  \jprl{100}, 121802 (2008). 
 
\bibitem{ref:Aubert:2007en} 
  B.~Aubert {\it et al.} (\babar\ Collaboration), 
  \jprd{78}, 011105 (2008). 
 
\bibitem{ref:Aubert:2009ck} 
  B.~Aubert {\it et al.} (\babar\ Collaboration), 
  \jprd{80}, 071103 (2009). 
 
\bibitem{ref:Aubert:2008zh} 
  B.~Aubert {\it et al.} (\babar\ Collaboration), 
  \jprl{103}, 211801 (2009). 
 
 
 
\bibitem{ref:Aubert:2001tu} 
  B.~Aubert {\it et al.} (\babar\ Collaboration), 
  \nima{479}, 1 (2002). 
 
 
\bibitem{ref:chargeconj} Reference to the charge-conjugate state is implied here and throughout the text unless otherwise stated.        
 
 
\bibitem{ref:babar_gamma_dalitz2008} B.~Aubert {\em et al.} (\babar\ Collaboration), \jprd{78}, 034023 (2008).  
 
 
\bibitem{ref:Asner:2005sz} 
  D.~M.~Asner {\it et al.} (CLEO Collaboration), 
  \jprd{72}, 012001 (2005). 
 
\bibitem{ref:Abe:2007rd} 
  L.~M.~Zhang {\it et al.} (Belle Collaboration), 
  \jprl{99}, 131803 (2007). 
 
 
 
 
 
\bibitem{ref:pdg2008} C.~Amsler {\it et al.} (Particle Data Group), \plb{667}, 1 (2008).    
 
\bibitem{ref:epaps} 
                    See supplementary material for additional mixing fit results and figures, and break-down of systematic uncertainties. 
 
\bibitem{ref:RevDalitzPlotFormalism} See review on Dalitz plot analysis formalism in~\cite{ref:pdg2008}. 
 
 
\bibitem{ref:AS} V.~V.~Anisovich and A.~V.~Sarantsev, \epj{A16}, 229 (2003). 
 
\bibitem{ref:LASS} D.~Aston {\em et al.}, \npb{296}, 493 (1988); W.~Dunwoodie, private communication. 
 
 
\bibitem{ref:Kmatrix-Pvector} I.~J.~R.~Aitchison, \npa{189}, 417 (1972). 
 
 
\bibitem{ref:CLEOc-KpiPhase}  J.~L.~Rosner {\em et al.} (CLEO Collaboration), \jprl{100}, 221801 (2008). 
 
\bibitem{ref:hfagFPCP10} A.~Schwartz {\it et al.}  (Heavy Flavor Averaging Group), 
\url{http://www.slac.stanford.edu/xorg/hfag/charm/index.html} (2010). 
 
 
 
 
 
 
 
 
 
 
 
 
 
 
 
 
 
 
 
 
\end{thebibliography}
\end{document}